\documentclass{aa}

\usepackage{graphicx}
\usepackage{dcolumn}
\usepackage{bm}
\usepackage{aas_macros}
\usepackage{hyperref}

\usepackage{txfonts}
\usepackage{multirow}

\usepackage{graphicx}
\usepackage{xcolor}
\usepackage{soul}

\usepackage{lipsum}

\usepackage{esvect}

\begin{document}

\title{PyGRO: a Python Integrator for General Relativistic Orbits}
\titlerunning{PyGRO}
\author{Riccardo Della Monica\thanks{Email: rdellamonica@usal.es}}
\authorrunning{R.~Della~Monica}
\institute{Departamento de Física Fundamental y Matemáticas, Universidad de Salamanca, Patio de Escuelas 1, Salamanca, 37008, Spain}

\date{\today}

\abstract
{Advancement in recent years in the field of experimental gravitation has allowed to test the equivalence principle in regimes that were previously unexplored, allowing for unprecedented verifications of general relativity and also enabling tests of alternative theories of gravity. We introduce a new computational tool, PyGRO, with the aim of integrating numerically the geodesic equations for the trajectories of massive and massless test particles in any analytic four-dimensional space-time. The result is a modern, fast, open-source, highly customizable and user-friendly Python package to perform the numerical integration of the geodesic equations. Combining symbolic and numerical calculations, PyGRO offers a variety of methods to obtain fully relativistic orbits with minimal intervention by the user, working in full generality with any user-given symbolic expression of the space-time metric tensor. We have tested PyGRO in an array of different scenarios, validating the methodology employed by successfully reproducing classical results from general relativity, which we report in this article.}

\maketitle

\section{Introduction}

The equivalence principle lies at the heart of general relativity and has been pivotal in shaping our understanding of gravity \citep{Dicke1964}. Historically, it provided the conceptual foundation for the formulation of Einstein's theory \citep{Einstein1915a} and has served as the basis for classical tests of gravity. These classical tests -- namely, the deflection of light by the Sun \citep{Eddington1920}, the precession of Mercury's perihelion \citep{Einstein1915b}, and the gravitational redshift \citep{Pound1960} -- stand as cornerstones of experimental gravitation, providing early confirmations of general relativity. A direct consequence of the equivalence principle is the geodesic nature of motion, which dictates that freely falling test particles, whether massive or massless, follow trajectories determined solely by the underlying geometry of space-time.

This fundamental aspect of relativistic dynamics has allowed experimental gravitation to advance significantly over recent decades, leading to numerous groundbreaking discoveries and increasingly precise tests of the equivalence principle at different scales \citep{Will2014}. Notable modern developments include missions such as Gravity Probe A \citep{Vessot1980} and Gravity Probe B \citep{Everitt2011}, which measured gravitational time dilation and the frame-dragging effect with unheard-of precision, along with the detection of strong gravitational lensing \citep{Bernstein1993} and the confirmation that black holes are actual astrophysical entities \citep{Bolton1972,Webster1972}. Additionally, the recent advent of gravitational wave astronomy \citep{Abbott2016} marked the birth of a completely new avenue for experimental gravitation.

One of the most exciting developments in the past thirty years is the detailed study of the Galactic center and its supermassive black hole, Sagittarius A* (Sgr A*) \citep{DeLaurentis2023}. Observations have revealed the existence of a large population of young stars, called the S-stars, moving on tight, highly eccentric orbits around Sgr A* \citep{Eckart1996,Ghez1998}. These stars provided the first definitive evidence for the presence of a supermassive black hole at the Galactic center \citep{Genzel2010}. Over time, precise measurements of their orbits have enabled direct tests of relativistic effects in a strong gravitational field \citep{Hees2017}. The star S2, in particular, has been crucial in this effort, with its periapsis passage in 2018 confirming the predicted gravitational redshift and orbital precession as described by general relativity \citep{Do2019,GravityCollaboration2018,GravityCollaboration2020}.

The successful imaging of supermassive black holes M87* \citep{EHT2019} and Sgr A* \citep{EHT2022} by the Event Horizon Telescope (EHT) represents another milestone in experimental gravitation, offering a direct detection of the shadow of a black hole, thus providing a strong-field test of gravitational lensing by a compact object \citep{Psaltis2020}. Moreover, the recent detection of polarized emission on the event-horizon scales of both M87* and Sgr A* \citep{EventHorizonTelescope2021a, EventHorizonTelescope2024} allowed to probe the structure of magnetic fields and the plasma properties near supermassive black holes \citep{EventHorizonTelescope2021b}.

Looking ahead, future experiments promise to push the boundaries of experimental gravitation even further. The next-generation EHT \citep[ngEHT]{Johnson2023} will enable the direct observation of black hole photon rings, while instruments such as GRAVITY+ \citep{GravityPlus2022} at the Very Large Telescope (VLT), MICADO \citep{Sturm2024} at the Extremely Large Telescope (ELT), and the Thirty Meter Telescope \citep{Skidmore2015} in Hawaii will allow us to resolve even closer stars in the Galactic center, diving deeper into the relativistic regime. Moreover, the potential discovery of pulsars orbiting a supermassive black hole by facilities like the Square Kilometre Array \citep[SKA]{Keane2015}, the Five-hundred-meter Aperture Spherical Telescope \citep[FAST]{Nan2011} and the next-generation Very Large Array \citep[ngVLA]{DiFrancesco2019} would open new avenues for precision tests of the equivalence principle and alternative theories of gravity \citep{DellaMonica2023e}.

\begin{figure*}
  \includegraphics[width=\textwidth]{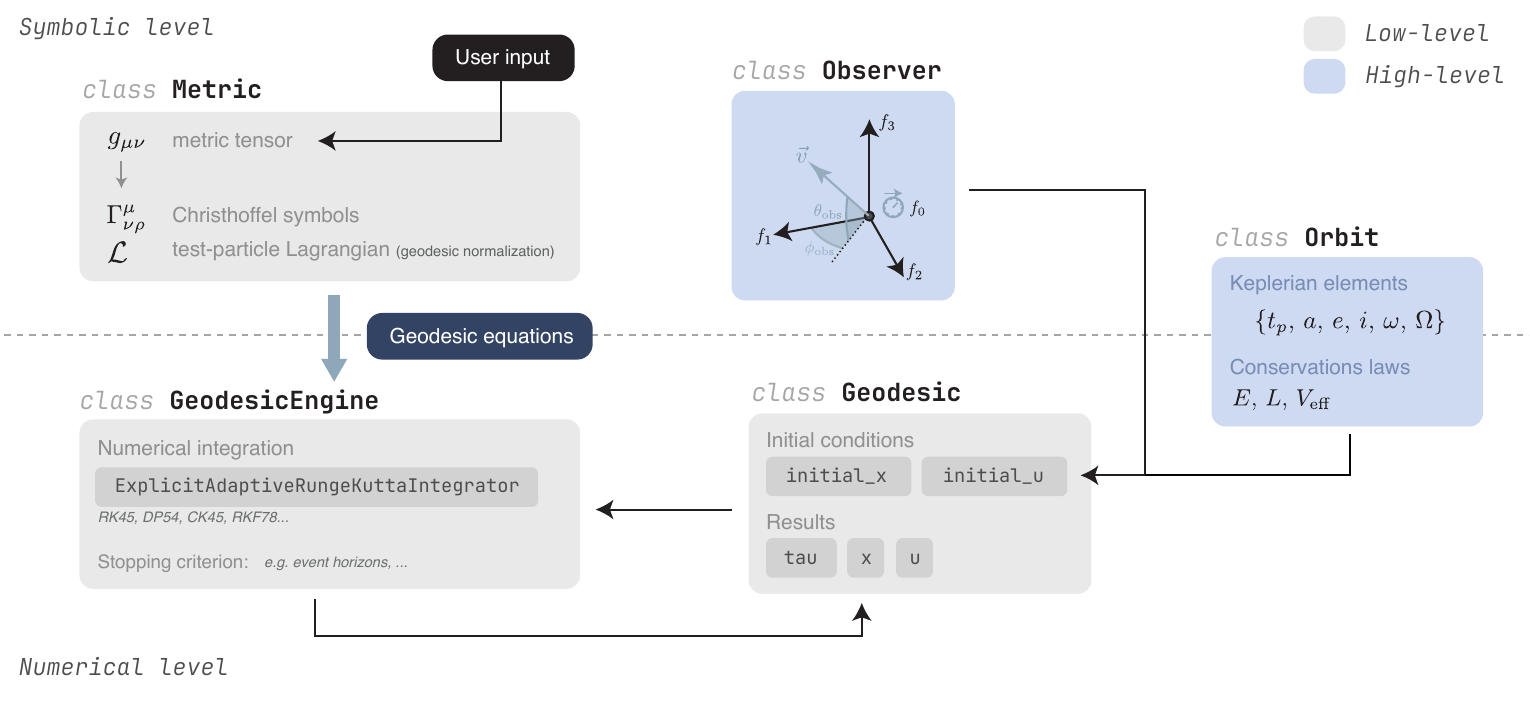}
  \caption{Schematic illustration of the code structure of PyGRO. The \texttt{Metric} object serves as the primary symbolic tool in PyGRO, computing all tensorial quantities required for deriving geodesic equations. The \texttt{GeodesicEngine}, on the other hand, is responsible for the numerical integration of these equations, acting as a worker that processes \texttt{Geodesic} objects. These objects encapsulate the geodesic type (e.g., null or time-like), properly normalized initial conditions (initial coordinates and initial tangent four-vector to the geodesic), and receive the integration results. Higher-level APIs in PyGRO, such as the \texttt{Observer} or the \texttt{Orbit}, comprising both symbolic and numerical calculations, provide a more physically intuitive approach to initializing geodesics.}
  \label{fig:code-structure}
\end{figure*}

In this context, we developed PyGRO, a Python integrator for General Relativistic Orbits. Building over numerous studies in the past years \citep{DeMartino2021,DellaMonica2022a,DellaMonica2022b,DellaMonica2023a,DellaMonica2023b,DellaMonica2023c,DellaMonica2023d,Cadoni2023,Fernandez2023,DeMoraLosada2025}, in which we have exploited observations and simulations of the S-stars at the Galactic center to constrain several alternative theories of gravity, black hole mimicker models and extended distribution of dark matter around Sgr A*, the code provides a computational tool designed to numerically integrate the geodesic equations for both massive and massless test particles in any analytic four-dimensional space-time. PyGRO is a modern, fast, open-source, highly customizable and user-friendly framework that combines symbolic and numerical methods to obtain fully relativistic orbits with minimal user intervention. In this article we will detail the structure and functionality of PyGRO (Section \ref{sec:code}), we will demonstrate its capabilities with various examples and benchmarks (Section \ref{sec:examples}), and we will highlight its potential as a powerful resource for studying relativistic dynamics in a wide range of astrophysically-relevant scenarios.

\section{The code structure}
\label{sec:code}

PyGRO is an open-source project distributed as a Python package\footnote{\url{https://pypi.org/project/pygro/}} and its documentation is publicly available on Github\footnote{\url{https://rdellamonica.github.io/pygro/index.html}}. The basic structure of PyGRO is shown in Fig. \ref{fig:code-structure}. PyGRO operates on both symbolic and numerical levels. The \texttt{Metric} engine (Section \ref{sec:metric}) represents the main symbolic tool within the code, dealing with the analytical computation of all the tensorial quantities required to compute the geodesic equations. The \texttt{GeodesicEngine} (Section \ref{sec:geodesic_engine}), on the other hand, is the main numerical class in PyGRO, performing the numerical integration of the geodesic equations. It acts as a worker to which integration jobs can be submitted in the form of \texttt{Geodesic} objects (Section \ref{sec:geodesic}), which store the geodesic type (e.g. null or time-like), the appropriately-normalized initial conditions (initial coordinates and initial tangent four-vector to the geodesic) and receives the results of the integration. Instead of assigning the initial conditions directly from the initial coordinates and initial tangent four-vector one can use higher-level APIs in PyGRO (Section \ref{sec:observer_orbit}), namely the \texttt{Observer} or the \texttt{Orbit} class in spherically-symmetric space-times, which allow a more physically intuitive assignment of initial data to geodesics.

\subsection{The \texttt{Metric} engine}
\label{sec:metric}

The \texttt{Metric} object serves as a fundamental tool for defining four-dimensional space-time metrics in PyGRO. Unlike other computational tools, like GYOTO \citep{Vincent2011}, that are tailored specifically for particular analytical solutions such as the Schwarzschild or Kerr metrics (or numerical metrics in the 3+1 formalism), the \texttt{Metric} class in PyGRO is designed with generality in mind. It allows users to specify an arbitrary coordinate system, $\{x^\mu\}_{\mu=0\dots3}$, and a symbolic expression of the metric tensor, $g_{\mu\nu}$ in the chosen chart, making it possible to study a wide range of space-time geometries. Conventionally, PyGRO adopts the space-time signature $(-,\,+,\,+,\,+)$.

The \texttt{Metric} class is built upon \texttt{sympy} \citep{Meurer2017} which allows, with minimal manual intervention, to derive from the user-input symbolic expression of the metric tensor all the essential quantities required to later integrate geodesics in that space-time. In fact, upon initialization, the \texttt{Metric} object computes the inverse metric $g^{\mu\nu}$, the Christoffel symbols of the metric, which in the Levi-Civita connection\footnote{While the Levi-Civita connection is built into PyGRO, it is fairly easy to implement any metric connection, from its symbolic expression.} take the form
\begin{equation}
  \Gamma^{\mu}_{\nu\rho} = \frac{1}{2}g^{\mu\sigma}(\partial_\nu g_{\sigma\rho}+\partial_\rho g_{\nu\sigma}-\partial_\sigma g_{\nu\rho}),
\end{equation}
from which the geodesic equations, that govern the motion of free-falling particles in the given space-time, are derived\footnote{It is fairly easy to introduce in PyGRO modifications to the geodesic equations, e.g. when considering interactions of test particles with a vector field which appear as non-zero term at the right-hand-side of Eq. \eqref{eq:geodesic_equation}, as done in \citet{DellaMonica2022b}.}:
\begin{equation}
  \ddot{x}^\mu + \Gamma^{\mu}_{\nu\rho}\dot{x}^\mu\dot{x}^\nu = 0,
  \label{eq:geodesic_equation}
\end{equation}
where dots represent derivatives with respect to an affine parameter $\tau$ on the geodesic (coinciding with the proper time for the time-like case) and the functions $\{x^\mu(\tau)\}_{\mu=0\dots 3}$ are its unknown space-time coordinates as a function of $\tau$.
Moreover, PyGRO symbolically derives helper functions that allow to normalize the initial tangent four-vector to the geodesic, defining its causal character:
\begin{equation}
  g_{\mu\nu}\dot{x}^\mu\dot{x}^\nu = \left\{
    \begin{array}{ll}
      -1& \qquad\textrm{time-like}\\
      0& \qquad\textrm{null}
    \end{array}\right..
    \label{eq:geodesic_normalization}
  \end{equation}
  This allows studying the motion of both massive particles and light rays in PyGRO. More specifically, these helper functions allow to compute one of the components of the initial tangent four-vector from given values of the other three components, solving Eq. \eqref{eq:geodesic_normalization} for the unknown one, once the causal character of the geodesic has been defined.

  The \texttt{Metric} object also allows computing the test-particle Lagrangian
  \begin{equation}
    \mathcal{L}=\frac{1}{2}g_{\mu\nu}\dot{x}^\mu\dot{x}^\nu,
    \label{eq:lagrangian}
  \end{equation}
  which is particularly useful to derive conserved quantities that are needed for higher level APIs, like the \texttt{Orbit} class.

  A key advantage of the symbolic approach presented in this section is its versatility: users can define arbitrary metrics without requiring explicit derivations of Christoffel symbols or other geometric quantities. However, this flexibility comes at a computational cost compared to hardcoded implementations in lower-level languages such as C, where the metric and its derived quantities are precomputed and optimized for specific cases. In PyGRO, especially for algebraically complicated metric tensors, symbolic differentiation and matrix operations introduce an overhead. While PyGRO prioritizes user-friendliness and generality over raw computational speed, its performance remains suitable for a broad range of applications. These aspects are all quantified in Section \ref{sec:performances}.

  \subsection{The \texttt{GeodesicEngine}}
  \label{sec:geodesic_engine}

  The \texttt{GeodesicEngine} class in PyGRO is the primary numerical tool designed to integrate geodesic equations within a specified space-time metric. By linking to an initialized \texttt{Metric} object, the \texttt{GeodesicEngine} utilizes the pre-computed symbolic quantities to perform numerical integrations of geodesics. Users can specify the symbolic backend used for the integration. The default, \emph{autowrap}, converts the symbolic expressions of the geodesic equations into pre-compiled C functions, which gives a $\sim2\times$ boost in performance during integrations. Alternatively, the \emph{lambdify} backend converts expressions into Python callables, which may be preferable in environments lacking a C compiler or when the metric is defined upon auxiliary Python functions (lacking a closed-form analytical expression). Additionally, users can select from various explicit embedded adaptive-step-size ordinary differential equation integrators of the Runge-Kutta family, such as Runge-Kutta-Fehlberg4(5), Dormand-Prince5(4), Cash-Karp4(5), and Runge-Kutta-Fehlberg7(8) \citep{Press1992}, each offering different orders of accuracy and truncation error estimation methods. Users can configure parameters like maximum and minimum step sizes and the accuracy (AG) and precision (PG) goal by which absolute and relative tolerances
  \begin{align}
    \textrm{tol}_a &= 10^{-\textrm{AG}}& \textrm{tol}_r &= 10^{-\textrm{PG}}
  \end{align}
  are defined. At each step the integrator adapts the step size in order to make the truncation error $\varepsilon$, estimated from the embedded Runge-Kutta formulas, be below a tolerance threshold defined as
  \begin{equation}
    \textrm{tol} = \textrm{tol}_a + \textrm{tol}_r ||y||,
    \label{eq:tolerances}
  \end{equation}
  being $||y||$ the norm of the current state vector. In the future we also plan to implement implicit schemes of integration and symplectic integrators.

  An important feature of the \texttt{GeodesicEngine} is the possibility to handle stopping criteria during integration. Users can define conditions under which the integration should halt, such as when a particle reaches a specific region of space-time. This is particularly useful, for instance, in the Schwarzschild metric, in which one might set a stopping criterion at the event horizon ($r = 2M$).

  \subsection{The low level geodesic API: \texttt{Geodesic}}
  \label{sec:geodesic}

  The \texttt{Geodesic} class in PyGRO represents the core data structure for describing geodesics in a given space-time. Each \texttt{Geodesic} instance stores essential information, such as the type of geodesic (time-like or null), the initial position and tangent four-vector to the geodesic, normalized to ensure consistency with the geodesic causal character, and stores the results of the integration from the \texttt{GeodesicEngine}.

  The \texttt{Geodesic} class represents a low-level API for the geodesic integration. Initial data, in fact, must be assigned directly in terms of the initial coordinates in the given chart and the components of the geodesic's tangent four-vector in the coordinate basis (which, in general, lack a direct physical interpretation). For this specific purpose, the \texttt{Geodesic} class makes use of the helper symbolic functions computed by the \texttt{Metric} object to compute one of the components of the initial tangent four-vector starting from the values assigned to the other three, solving Eq. \eqref{eq:geodesic_normalization} for the unknown component.

  \subsection{High level APIs: \texttt{Observer} and \texttt{Orbit}}
  \label{sec:observer_orbit}

  Assigning initial conditions that have a meaningful physical meaning is important to compute geodesic trajectories that are relevant for astrophysical applications of PyGRO. Directly assigning the initial values of the space-time coordinates and initial tangent four-vector to \texttt{Geodesic} objects can be avoided making use of higher-level APIs in PyGRO that allow to specify positions and velocities tied to the reference frame of a given physical observer in the space-time (Section \ref{sec:observer}), or that have a direct connection with the classical Keplerian parametrization of orbits in Newtonian mechanics (Section \ref{sec:orbit}).

  \subsubsection{Defining physical observers in PyGRO}
  \label{sec:observer}

  \begin{figure}
    \includegraphics[width=\columnwidth]{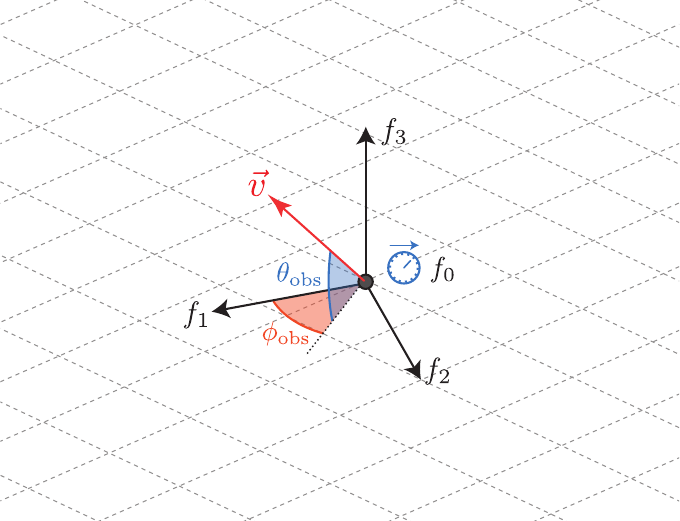}
    \caption{Illustration of the reference frame carried by a physical observer in PyGRO. Angles of observations in the observer's reference frame are defined as standard longitude, $\phi_\textrm{obs}$ and latitude $\theta_\textrm{obs}$ with origin on the ${f}_1$ direction and reference plane identified by ${f}_1$ and ${f}_2$. Helper functions built into the \texttt{Observer} class allow to fire geodesics with any given pair of angles, defined in this way, from any of the frame vectors.}
    \label{fig:observer}
  \end{figure}

  The \texttt{Observer} class in PyGRO provides a high-level interface for defining observers in any given space-time and assigning initial data to geodesics in a physically meaningful way. This abstraction is particularly useful in astrophysical settings, where one wants to compute trajectories of freely-falling objects and photons as observed by a specific observer (e.g. in a ray-tracing scenario to image a black hole shadow, to compute the astrometric positions of stars around a massive black hole as seen by a given observer, or to perform pulsar timing).

  The approach to general relativity that relies on observer frames is known as the tetrad formalism (or Cartan formalism) \citep{Straumann2013}. In this framework, one constructs frames -- also referred to as vierbeins -- at each point on the curved manifold, i.e., constructing a frame bundle over the manifold. These frames consist of four orthonormal vectors denoted as $\{ f^\alpha_a \}$, where Greek indices refer to the components of the frame vectors in the chart coordinates on the manifold, while Roman indices label each frame vector. The set $\{ f^\alpha_a \}$ contains one time-like vector ($f^\alpha_0$) and three space-like vectors ($f^\alpha_i$, $i = 1,\,2,\,3$), representing, respectively, the observer's time axis (which coincides with the physical observer's four-velocity) and a set of orthonormal spatial axes forming a spatial tetrad at that point. The observer uses these axes to define tangent vectors to the manifold and project them conveniently into its own frame. In this sense, the tetrad provides a local basis for the tangent space at the observer's location, enabling the projection of physical quantities from the global manifold coordinates into the local reference frame of the observer.

  Similarly, one can define a co-frame $\{f_\beta^b\}$, which forms the basis for tangent 1-forms at the observer's location. This can be generalized, allowing one to use the defined frame and co-frame basis to represent any tensorial quantity in the tangent or co-tangent bundle (or any product of these). For instance, the metric tensor can be expressed in the co-frame as
  \begin{equation}
    g = -f^0 \otimes f^0 + \sum_i f^i \otimes f^i,
    \label{eq:metric-tetrad}
  \end{equation}
  which allows to obtain relations between the local frame metric and the space-time metric:
  \begin{align}
    f_\alpha^a f_\beta^b \eta_{ab} &= g_{\alpha\beta}, \\
    f^\alpha_a f^\beta_b g_{\alpha\beta} &= \eta_{ab}.
  \end{align}
  In other words, in the local observer's frame, the metric reduces to the ordinary Minkowski metric, which reflects the local flatness property of general relativity.

  In PyGRO the \texttt{Observer} class is built starting from a symbolic expression of the frame (co-frame) vectors in the coordinate basis of the tangent (co-tangent) space to the manifold. Upon definition, the code autonomously computes symbolic expressions of the transformation matrices from the observer's reference frame to the space-time coordinate basis and vice-versa.

  This allows to place an observer at a specific event in space-time, carrying a local system of orthonormal axes on which one can project physical quantities, e.g.  four-velocities of test particles. More specifically, one can define a spatial vector $\vv{v}$ in this system of axes, whose direction is identified by polar coordinates $(\theta_\textrm{obs},\,\phi_\textrm{obs})$ and with a given magnitude $|\vv{v}|$ (see Fig. \ref{fig:observer}). By projecting this vector on the observer's frame, one can derive in the coordinate basis the components of the corresponding four-vector $v^\mu$ whose spatial part coincides with $\vv{v}$, once a given causal character for $v^\mu$ has been chosen (the temporal component, $v^0$, is fixed by the normalization condition in Eq. \eqref{eq:geodesic_normalization} with $\dot{x}^\mu$ replaced by $v^\mu$). In case the desired four-vector is null, the magnitude of $\vv{v}$ is automatically normalized to unity. The four-vector obtained with this procedure can then be used as initial tangent four-vector for a geodesic starting at the observer's location, which will correspond to a test particle which, according to the observer, has a physical velocity $\vv{v}$ in its reference frame.

  \subsubsection{Quasi-elliptic orbits in spherically symmetric space-times}
  \label{sec:orbit}

  \begin{figure}
    \includegraphics[width=\columnwidth]{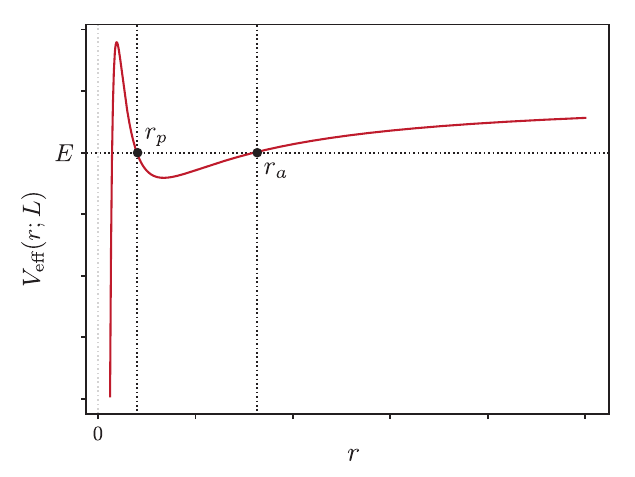}
    \caption{Effective potential for given values of $E$ and $L$ in the Schwarzschild space-time. A choice of $E$ and $L$ uniquely identifies radial turning points, i.e. values of the radial coordinates at which the radial velocity $\dot{r} = 0$. These points are identified by $r_p$, the pericenter, and $r_a$, the apocenter.}
    \label{fig:effective-potential}
  \end{figure}

  \begin{figure}
    \includegraphics[width=\columnwidth]{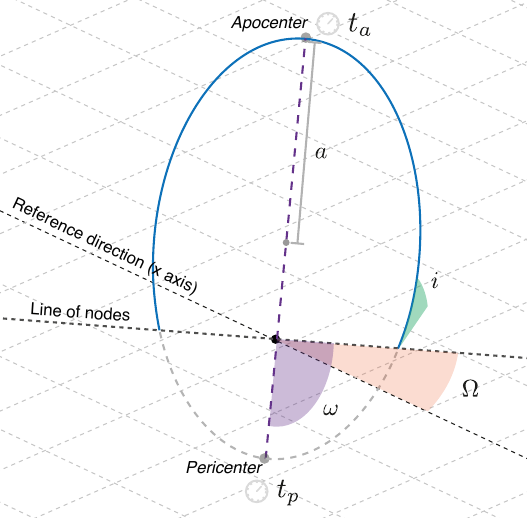}
    \caption{Illustration of the angular orbital elements for a Keplerian orbit. The orbital inclination $i$ represent the inclination of the orbital plane with respect to a reference plane (e.g. the plane of sky for a distant observer); the longitude of the ascending node $\Omega$ corresponds to the angle between a reference direction (in our case the $\phi=0$ line on the equatorial plane) and the line of nodes, namely the line in which the orbital plane intersects the reference plane; the argument of the pericenter $\omega$ is the angle, along the orbital plane between the line of nodes and the orbit's pericenter.}
    \label{fig:orbital-parameters}
  \end{figure}

  When the considered space-time is spherically symmetric \citep{Chandrasekhar1985}, assigning initial conditions to freely-falling massive test particles can be greatly simplified, enabling a parametrization of the orbit which is familiar to classical Keplerian parametrization of elliptical orbits in Newtonian celestial mechanics \citep{Taff1985}.

  In this section, we will assume that a chart of coordinates ($t$, $r$, $\theta$, $\phi$) has been chosen to express the metric tensor, with $t$ a coordinate-time, $r$ a radial coordinate (for example, but not necessarily, the aerial radius of the Schwarzschild coordinate system), and $\theta$ and $\phi$ are the usual angular coordinates on the 2-spheres. In this setting, the geodesic equations in Eq. \eqref{eq:geodesic_equation} are a system of four second-order differential equations for the unknown functions ($t(\tau)$, $r(\tau)$, $\theta(\tau)$, $\phi(\tau)$) that describe the space-time trajectory of a test particle. A solution is uniquely determined once second-order initial conditions are assigned. Namely, one has to specify at a given proper time $\tau_0$ the initial space-time position of the test particle ($t(\tau_0)$, $r(\tau_0)$, $\theta(\tau_0)$, $\phi(\tau_0)$) and space-time four-velocity ($\dot{t}(\tau_0)$, $\dot{r}(\tau_0)$, $\dot{\theta}(\tau_0)$, $\dot{\phi}(\tau_0)$). Importantly, due to the spherical symmetry, one can always perform appropriate rotations to make the trajectory lie on the equatorial plane for the whole duration of the motion. This reduces the number of free parameters, since $\theta = \pi/2$ and $\dot{\theta} = 0$ are assumed implicitly. Second, since the metric components (and thus the geodesic equations) do not explicitly depend on the time coordinate $t$, the value $t(\tau_0)$ can be chosen arbitrarily. Finally, as mentioned above, the normalization condition of the four-velocity for massive test particles, in Eq. \eqref{eq:geodesic_normalization}, represents a constraint on the initial data, thus allowing to derive one of the components of the four-velocity as a function of the others. A complete set of initial data that uniquely identifies a trajectory on the equatorial plane of a spherically symmetric space-time can thus be ($r(\tau_0)$, $\phi(\tau_0)$, $\dot{t}(\tau_0)$, $\dot{\phi}(\tau_0)$), with $\dot{r}(\tau_0)$ fixed by the choice of $\dot{t}(\tau_0)$ and $\dot{\phi}(\tau_0)$ through the normalization condition. Instead of fixing the initial conditions directly in terms of these components of the four-velocity, we introduce a different parameterization in terms of constants of motion. In particular, in spherically symmetric space-times from the relativistic test-particle Lagrangian in Eq. \eqref{eq:lagrangian} (which is by itself a constant of motion, since, modulo the factor $1/2$, it coincides with the norm of the particle's four-velocity) one can always define two additional constants of motion,
  \begin{align}
    E &= -\frac{\partial \mathcal{L}}{\partial \dot{t}},\label{eq:specific-energy}\\
    L &= \frac{\partial \mathcal{L}}{\partial \dot{\phi}},
    \label{eq:angular-momentum}
  \end{align}
  corresponding to the specific energy and angular momentum of the test particle. These constants allow to reduce the problem of describing the radial motion on the equatorial plane to the equation
  \begin{equation}
    \dot{r}^2 = E - V_\textrm{eff}(r; L),
    \label{eq:dotr2}
  \end{equation}
  where we have introduced the effective potential $V_\textrm{eff}$, parametrized by the orbital angular momentum $L$, which only depends functionally on the radial coordinate $r$. A choice of $E$ and $L$ uniquely fixes the evolution of the radial coordinate and, more specifically, defines the radial turning points, i.e. points where $\dot{r} = 0$. In the case of quasi-elliptic orbits, we call the two radial turning points the pericenter (identified by the radial coordinate $r_p$) and the apocenter (identified by $r_a$). These are shown in Fig. \ref{fig:effective-potential} for an example orbit in the Schwarzschild space-time.

  More conveniently, one can introduce two orbital parameters, namely the eccentricity $e$ and the semi-major axis $a$, defined implicitly by
  \begin{align}
    r_p &= a(1-e)   \\
    r_a &= a(1+e),
    \label{eq:radial-turning-points}
  \end{align}
  whose values are uniquely identified by a choice of $E$ and $L$. We can invert this mapping so that unique values for $E$ and $L$ are assigned upon fixing a value for $a$ and $e$. This inverse map can be derived analytically for the Schwarzschild space-time. However, to preserve the generality of the approach to any spherically-symmetric space-time, in PyGRO the mapping from pairs ($a$, $e$) to pairs ($E$, $L$) is computed numerically within the \texttt{Orbit} class. Once the values of $E$ and $L$ are known, one can invert equations Eqs. \eqref{eq:specific-energy} and \eqref{eq:angular-momentum} to obtain $\dot{t}$ and $\dot{\phi}$.

  In the flat space-time limit, the orbital parameters that we have introduced perfectly match the concept of semi-major axis and eccentricity classically defined in Newtonian celestial mechanics for Keplerian elliptical orbits. In the general relativistic case, these parameters depend on the particular choice of coordinates used (for example, in the harmonic gauge of post-Newtonian mechanics, these parameters have a slightly different definition and physical interpretation, e.g. \citeauthor{Damour1985}, \citeyear{Damour1985}), and their physical meaning is not uniquely defined in the strong-field regime. As such, they do not have a direct physical meaning but only serve as a familiar and useful parametrization of initial data.

  We can now choose whether to start the integration at the apocenter or pericenter and fix the rest of the initial conditions from there. For example, if we call $t_p$ the time of pericenter passage, we have that a set of initial data are
  \begin{align}
    t(\tau_0) &= t_p, \label{eq:initial-conditions-1}\\
    r(\tau_0) &= a(1-e), \\
    \phi(\tau_0) &= 0, \\
    \dot{r}(\tau_0) &= 0,\label{eq:initial-conditions-4}
  \end{align}
  and $\dot{t}(\tau_0)$ and $\dot{\phi}(\tau_0)$ are derived using the procedure that we have just explained. Alternatively, one can fix initial conditions at apocenter, in whose case one would have
  \begin{align}
    t(\tau_0) &= t_a, \label{eq:initial-conditions-apo-1}\\
    r(\tau_0) &= a(1+e), \\
    \phi(\tau_0) &= \pi, \\
    \dot{r}(\tau_0) &= 0,\label{eq:initial-conditions-apo-4}
  \end{align}
  where we call $t_a$ the time of passage at apocenter. Note that the choice of $\phi(\tau_0)=0$ at pericenter, or alternatively $\phi(\tau_0)=\pi$ at apocenter, comes from the analogy between the azimuthal coordinate $\phi$ and the true anomaly of the orbit. This, of course, is only valid at the initial time since the relativistic orbital precession (which will be described in more details in Section \ref{sec:precession}) causes the two quantities to eventually deviate one another with a constant rate.

  The procedure described above completely fixes the initial conditions on the equatorial plane and allows to integrate the geodesic equations numerically. To bring the orbit outside the equatorial plane, one can introduce the standard angular Keplerian parameters (see Fig. \ref{fig:orbital-parameters}). In particular: the inclination ($i$) measures how much the orbital plane is inclined with respect to the equatorial plane of the spherically symmetric space-time considered. The two planes intersect on a line known as the line of nodes. The orbiting object crosses this line twice over one period. The point where it cuts the line from below the plane is known as the ascending node, while the point where it cuts it from above is known as the descending node; the longitude of the ascending node ($\Omega$) is the angle between the $x$-axis (i.e., the one identified by $\theta = \pi/2$ and $\phi = 0$) and the ascending node; the argument of the pericenter ($\omega$) is the angle between the line of nodes and the semi-major axis of the orbit, specifically its pericenter, over the orbital plane. These angles correspond to three subsequent rotations, that are applied to both the initial position and velocity before integrating the geodesic, that bring the orbit into the desired reference frame:
  (i) a rotation around the $z$-axis by an angle $-\omega$ (the minus sign is due to the fact that by definition, $\omega$ points towards $x$ and not from it); (ii) a rotation by an angle $-i$ around the new $x$-axis, corresponding to the line of nodes; (iii) a rotation around the $z$-axis by an angle $-\Omega$. It is important to note that in spherically symmetric spacetimes, the existence of an SO(3) isometry subgroup ensures that rotations can always be defined in a covariant manner. This allows one to orient a given geodesic into a preferred plane, often the equatorial plane, without loss of generality, as done for the covariant definition of the equatorial plane motion. However, the angular orbital elements such as $i$,  $\omega$ and $\Omega$, introduced above, are defined with respect to a chosen reference plane and thus depend on the coordinate system. While useful for comparisons with Newtonian orbital elements, they are not intrinsic to the geodesic itself.

  The methodology that we have explained in this section, requires a combination of symbolic and numerical calculations that are implemented in the \texttt{Orbit} class. This acts as a wrapper to the \texttt{Geodesic} class, for the time-like case, and allows to integrate an orbit once values for the Keplerian orbital parameter ($t_p$, $a$, $e$, $i$, $\Omega$, $\omega$)\footnote{Or, alternatively, ($t_a$, $a$, $e$, $i$, $\Omega$, $\omega$), being $t_a$ the time of apocenter passage. In this case, the initial conditions will be assigned at $t(\tau_0) = t_a$, $r(\tau_0) = a(1+e)$, $\phi(\tau_0) = \pi$.} are assigned.

  \section{Examples and benchmarks}
  \label{sec:examples}

  \begin{figure}
    \includegraphics[width=\columnwidth]{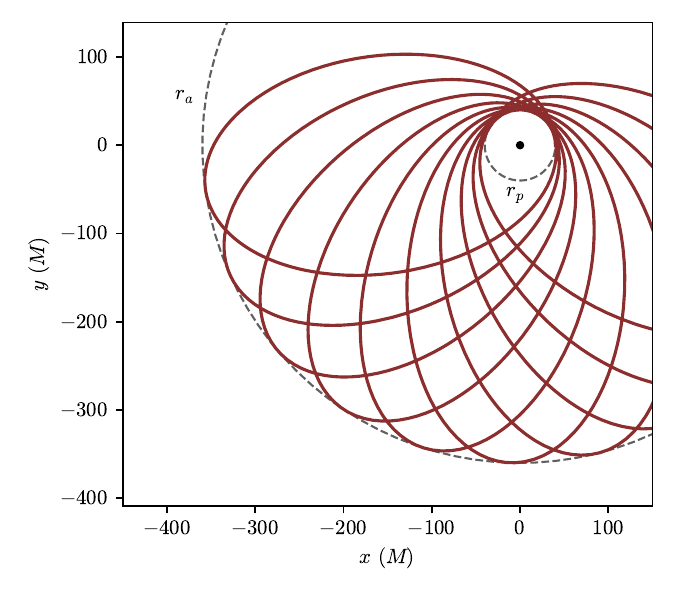}
    \caption{Example orbit with $e = 0.8$ and $a = 200M$ integrated with PyGRO. The integration has been carried on for 10 full revolutions around the central object. The trajectory is not a closed ellipse. Orbital precession makes the orbit's pericenter advance, while the radial coordinate periodically oscillates between the pericenter and apocenter, Eqs. \ref{eq:radial-turning-points}, here indicated by the dashed circles. This example orbit is labeled as (1) in Table \ref{tab:runtimes}.}
    \label{fig:orbital-precession-1}
  \end{figure}

  \begin{figure}
    \includegraphics[width=\columnwidth]{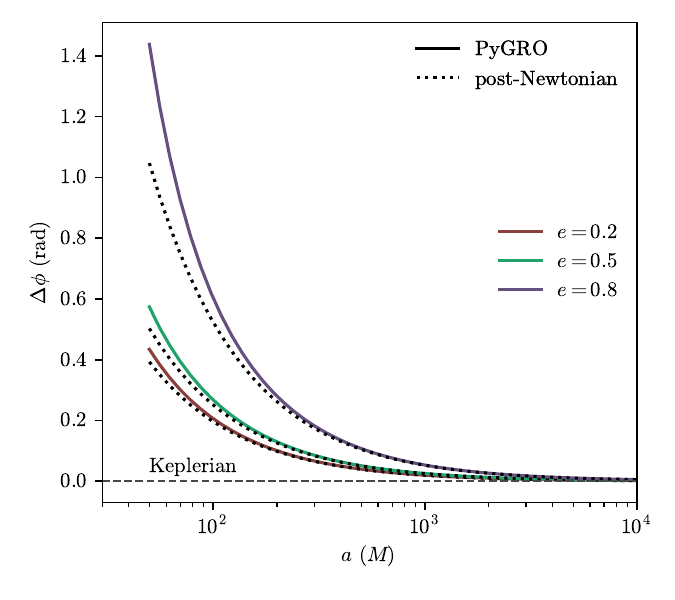}
    \caption{The rate of orbital precession for orbits with $a\in[50, 10000]$ and for three values of the eccentricity, $e=0.2$, $0.5$ and $0.8$. The solid lines correspond to the rate of precession computed with PyGRO, while the dotted lines represent the post-Newtonian prediction for the same orbits, computed with the formula in  Eq. \eqref{eq:precession}. PyGRO correctly recovers the values of the predicted rate of precession for systems in the weak field regime. Also, for systems that scan increasingly strong gravitational regimes (e.g. $a\lesssim100M$ in the low-eccentricity cases or even $a\lesssim1000M$ for higher eccentricities), the fully-relativistic computations in PyGRO allow to go beyond the first-order post-Newtonian prediction and correctly account for an additional contribution to the periastron advance.}
    \label{fig:orbital-precession-2}
  \end{figure}

  In this section we will analyze a series of example applications of PyGRO that will work as a benchmark of the methodology implemented in the code. Most of these examples will test a series of theoretical predicitons for the Schwarzschild space-time, here written assuming $G=c=1$,
  \begin{equation}
    ds^2 = -\left(1-\frac{2M}{r}\right)dt^2+\left(1-\frac{2M}{r}\right)^{-1}dr^2 + r^2d\Omega^2
    \label{eq:schwarzschild_metric}
  \end{equation}
  where $M$ is the mass of the central object, $r$ is the aerial radius and $d\Omega^2=d\theta^2+\sin^2\theta d\phi^2$ is the solid angle element. The correct recovery of theoretical predictions for the Schwarzschild metric effectively validates PyGRO and works as a benchmark for the methodology that we have discussed.

  The code to reproduce all the examples shown in this section can be found in the \href{https://rdellamonica.github.io/pygro/index.html}{example notebooks} in PyGRO's official documentation.

  \subsection{Orbital precession of massive particles}
  \label{sec:precession}

  Here, we illustrate how in PyGRO one can accurately reproduce the expected orbital motion for bound time-like orbits and also determine numerically the rate of orbital precession in the strong-field regime.

  Let us consider eccentric orbits on the equatorial plane of the Schwarzschild black hole in Eq. \eqref{eq:schwarzschild_metric}. An orbit with semi-major axis $a$ and eccentricity $e$ is known to have a rate of orbital precession which, at first post-Newtonian order, can be quantified by \citep{Poisson2014}
  \begin{equation}
    \Delta\phi = \frac{6\pi GM}{ac^2(1 - e^2)},
    \label{eq:precession}
  \end{equation}
  per orbital period. This means that each complete oscillation on the radial direction between $r_a$ and $r_p$, corresponds to a complete turn in the azimuthal direciton, plus a small angle $\Delta\phi$ which thus corresponds to an advance of the periastron of the orbit.

  \begin{figure*}
    \includegraphics[width=\textwidth]{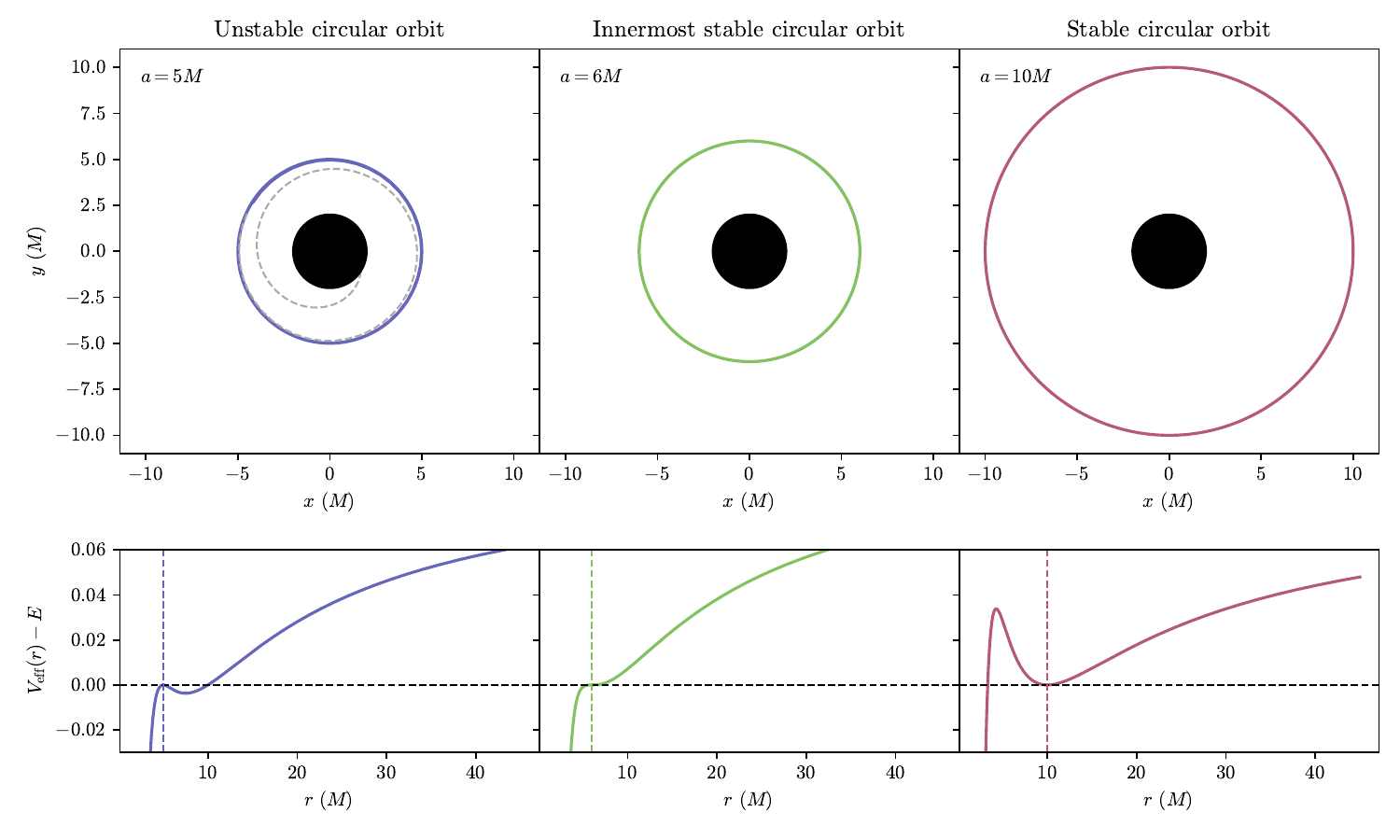}
    \caption{Circular orbits around a Schwarzschild black hole integrated with PyGRO. The top panels display the resulting orbits on the equatorial plane, with the black circle representing the central black hole. The bottom panels show the effective potential in Eq. \eqref{eq:dotr2} computed with PyGRO for the distinct orbital configurations. We analyze: (left panels) an unstable circular orbit with $a=5M$, thus lying on a maximum of the effective potential. The instability of the orbit manifests with the orbit falling into the black hole (gray dashed line) due to the accumulation of numerical errors when integrating for a long time interval. This example orbit is labeled as (2) in Table \ref{tab:runtimes}; (central panels) the case with $a=6M$ corresponding to the ISCO, and thus a saddle point in the effective potential; (right panels) a normal stable circular orbit which can be realized for all $a>6M$, corresponding to local minimums of the effective potential. This example orbit is labeled as (3) in Table \ref{tab:runtimes}.}
    \label{fig:circular_orbits}
  \end{figure*}

  In PyGRO, one can compute the rate of orbital precession by directly estimating the angle spanned by an orbit over one full cycle in the radial direction. To do so, first, we make use of the \texttt{Orbit} class to define orbits with a specific semi-major axis and eccentricity. An example orbit retrieved in the Schwarzschild metric with this methodology is shown in Fig. \ref{fig:orbital-precession-1}. Since we are fixing initial conditions like in Eqs. \eqref{eq:initial-conditions-1}-\eqref{eq:initial-conditions-4}, a full azimuthal cycle corresponds to an angle $\phi=2\pi$. Due to the orbital precession, however, this full cycle does not correspond to a full period in the radial direction, thus to a return to the pericenter radius $r_p$. This occurs at an angle $\phi=2\pi+\Delta\phi$, being $\Delta\phi$ the rate of orbital precession that we want to compute. On a practical level, this can be done by considering an interpolating function $\dot{r}=\dot{r}(\phi)$ built from the integrated geodesic and the subsequent application of a root finding algorithm, to look for the radial turning points (roots of $\dot{r}$) around $2\pi$. The difference of the angle obtained and $2\pi$ is the desired value of $\Delta\phi$.

  In Fig. \ref{fig:orbital-precession-2} we show the values of $\Delta\phi$ obtained with PyGRO for a large range of semi-major axes and for three eccentricities $e=0.2$, $0.5$ and $0.8$ in the case of a Schwarzschild black hole. We also compare our results with the post-Newtonian formula in Eq. \eqref{eq:precession}. As it appears, the post-Newtonian values is correctly recovered for orbits in a weak gravitational field (e.g. $a\gtrsim100M$ in the low-eccentricity cases or $a\gtrsim1000M$ for higher eccentricities), validating the accuracy of the relativistic orbital integration in PyGRO. For orbits in the strong field regime, on the other hand, the formula in Eq. \eqref{eq:precession} fails to describe accurately the rate of periastron advance, underestimating the actual value of $\Delta\phi$ that the code is able to compute. In this sense, PyGRO offers the advantage of not only being able to compute the relativistic rate of orbital precession directly on the fully-relativistic integrated geodesic, thus not having to resort to approximate formulas, but also the complete generality of the approach, which can be applied to any spherically symmetric space-time by simply defining a different \texttt{Metric} object.

  \subsection{Circular orbits}

  In this section, we study a different example of motion for massive particles on time-like geodesics in the Schwarzschild space-time: circular orbits.

  Orbits with zero eccentricity are characterized by $r_p=r_a=a$. For this reason, the initial condition search algorithm has to be slightly modified compared to the non-zero eccentricity case. We must now look for those values of specific energy $E$ and angular momentum $L$, Eqs. \eqref{eq:specific-energy}-\eqref{eq:angular-momentum}, that realize the conditions $\dot{r}^2(a)= 0$, and $V_\textrm{eff}'(a) =0$, with a prime denoting a derivative with respect to the radial coordinate. In other terms, we are looking for those value of energy and angular momentum that make so that the given value of the semi-major axis $a$ is a radial turning points, i.e. a root for Eq. \eqref{eq:dotr2}, and at the same time a stationary point for the effective potential. This procedure is built-into the \texttt{Orbit} class in PyGRO and no additional operation on the user's side is required to deal with circular orbits.

  To test the validity of this approach, we can study peculiar circular orbits that characterize the dynamical motion of massive test particles around a Schwarzschild black hole which present distinctive features not present in a Newtonian scenario. While in classical celestial mechanics it is possible to realize a circular orbital configuration for any value of $a$, around a point-like massive particle, in general relativity, and in particular in the Schwarzschild space-time, this is only possible for values of $a>3M$ \citep{Chandrasekhar1985}. No stable orbital configurations exist very close to the event horizon. Even in the regime where circular orbits do exist, they are not necessarily stable circular orbits. In fact, another known results from general relativity is that stable circular orbits only exist up to a minimum radius of $a = 6M$, denoted as the innermost stable circular orbit (ISCO). Orbits with $a>6M$ can be circular and are stable, meaning that they realize a minimum for the effective potential. Orbits with $3M<a<6M$ can still be circular but they are unstable, sitting on a maximum of the effective potential. Finally, the specific realization with $a = 6M$ corresponds to a saddle point of the effective potential, thus representing the separation between the two classes. In Fig. \ref{fig:circular_orbits} we display examples circular orbital configurations obtained with PyGRO for all the cases listed above, reproducing the results expected from the theory, along with the corresponding effective potential. In particular, along with a stable orbital configuration with $a=10M$ and the innermost stable circular orbit with $a = 6M$, both resulting perfectly circular from the integration, we show an unstable circular orbit with $a=5M$, which, after a large number of orbits falls into the central black hole due to the accumulation of numerical errors.\footnote{Clearly, one can always set the precision tolerances of the integrator to make the orbit stick to the circular configuration for the entire duration of the integration.}

  \subsection{Strong deflection of massless particles}
  \label{sec:photon_orbit}

  \begin{figure}
    \includegraphics[width=\columnwidth]{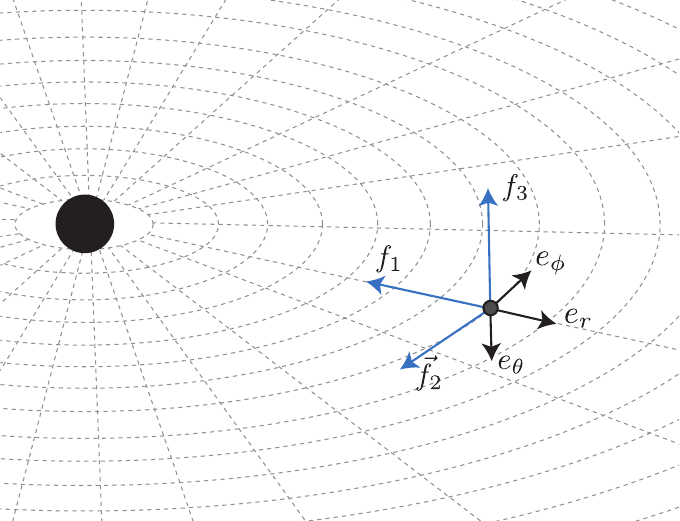}
    \caption{Illustration of reference frame for a stationary observer in the Schwarzschild case, defined as in Eqs. \eqref{eq:observer-frame-1}-\eqref{eq:observer-frame-4}. The observer (black dot) carries a left-handed system of axes, $f_1f_2f_3$,  with ${f}_1$ pointing radially towards the central black hole, ${f}_2$ pointing tangentially to $r=\textrm{const.}$ 2-spheres, opposite to the vector ${e}_\phi$ of the local coordinate basis, and ${f}_3$ orthonormal to both ${f}_1$ and ${f}_2$ (opposite to the vector ${e}_\theta$ of the local coordiante basis).}
    \label{fig:observer-schwarzschild}
  \end{figure}

  \begin{figure}
    \includegraphics[width=\columnwidth]{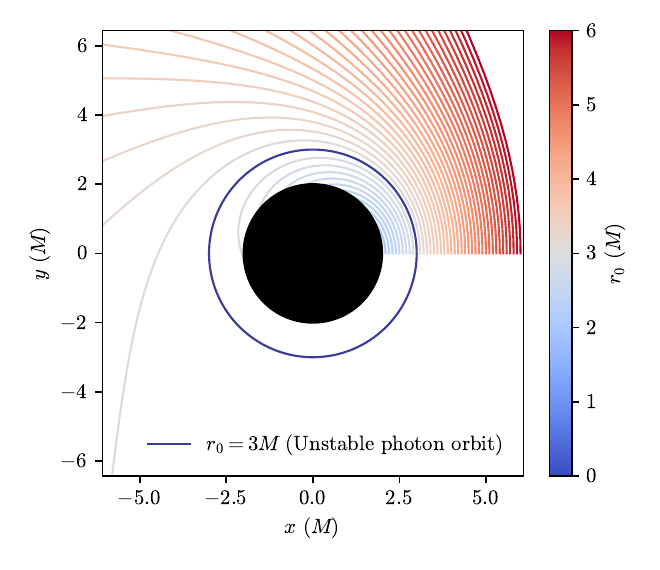}
    \caption{Photon trajectories obtained with PyGRO, fired by observers at positions $r_0\in[2,6]M$, around a Schwarzschild black hole with an initially tangential direction on the equatorial plane. Photons fired from below $r_0<3M$ end up in the horizon. Photons fired from $r_0>3M$ escape to infinity, after being strongly deflected by the central object (trajectories grazing closer to the $r_0=3M$ circumference experience a stronger lensing). Finally, for $r_0 = 3M$, the photon describes a circumference around the central black hole corresponding to the unstable photon orbit. In Table \ref{tab:runtimes} we label by (4), (5) and (6) the photons with $r_0=2.1M$, $3M$ and $6M$, respectively.}
    \label{fig:photon_orbits}
  \end{figure}

  In this section we analyzes the case of massless particles (e.g. photons) moving on null geodesics of the Schwarzschild space-time and being strongly deflected by the massive central object.

  Our goal is to recover in PyGRO one of the most striking predictions of general relativity, namely the existence of a circular (unstable) photon orbit for $r=3M$ (the so called photons sphere or last photon orbit) around a Schwarzschild black hole. This photon trajectory separates the classes of null geodesics that escape to infinity after being deflected by the central object and those captured by the black hole. This peculiar feature is what ultimately leads to the emergence of the shadow and photon rings in black hole imaging \citep{Luminet1979}.

  In this case, to assign initial conditions to the null geodesics we make use of the \texttt{Observer} class in PyGRO, introduced in Section \ref{sec:observer}. In particular, we will consider stationary observers in the Schwarzschild space-time, Eq. \eqref{eq:schwarzschild_metric}, with frame axes oriented along the coordinate lines of the Schwarzschild coordinate system. It is straightforward to see that such a system of observers can be defined by the following system of frame co-vectors:
  \begin{align}
    \allowdisplaybreaks
    f^0 &= \sqrt{1-\frac{2M}{r}}dt,& f^1 &= -\dfrac{1}{\sqrt{1-\frac{2M}{r}}}dr,\\
    f^2 &= -r\sin\theta d\phi,& f^3 &= -r d\theta,
  \end{align}
  which satisfies the relation in Eq. \eqref{eq:metric-tetrad}. This choice of frame co-vectors results in the frame vectors
  \begin{align}
    f_0 &= \dfrac{1}{\sqrt{1-\frac{2M}{r}}}\partial_t,&\label{eq:observer-frame-1} f_1 &= -\sqrt{1-\frac{2M}{r}}\partial_r,\\
    f_2 &= -\frac{1}{r\sin\theta} \partial_\phi,& f_3 &= -\frac{1}{r}\partial_\theta.\label{eq:observer-frame-4}
  \end{align}
  Essentially, the frame that we have defined describes an observer that is stationary (its time axis, $f_0$ is future-pointing and only directed in the temporal direction) and carries a system of $f_1f_2f_3$ left-handed axes with $f_1$ on the radial direction pointing inward towards the black hole, $f_2$ pointing tangentially in the longitudinal direction (opposite to the natural counter-clockwise direction of the angle $\phi$), and $f_3$ points ``upward'' in the latitudinal direction, opposite to the natural direction of the angle $\theta$. The geometrical configuration of the frame vectors for our observer in the Schwarzschild space-time is illustrated in Fig. \ref{fig:observer-schwarzschild}

  We now consider several observers defined by Eqs. \eqref{eq:observer-frame-1}-\eqref{eq:observer-frame-4} positioned at coordinates $r=r_0$, $\theta=\pi/2$ (i.e. lying on the equatorial plane) and $\phi=0$ with $r_0$ spanning from $2M$ (right on the horizon) to $6M$, and for each of them we fire a null geodesic from the $f_2$ axis, integrating it backward in time. Hence, we are considering photons reaching these observers from the direction of $f_2$ and tracing them back in time, as one usually does in raytracing applications. We obtain the photon paths depicted in Fig. \ref{fig:photon_orbits}. As one would expect from theoretical insights, photons reaching tangentially the observers at $r_0<3M$ hit the horizon. Photons reaching the observers at $r_0>3M$, on the other hand come from infinity, while at $r=3M$, the photon moves on a perfect circumference, revolving indefinitely around the central black hole.

  \subsection{Reconstructing astronomical observables: the S2 star in the Galactic center}

  In this example, we will study how one can use the \texttt{Orbit} class in PyGRO and the results of the numerical integration of the geodesic equations to obtain astronomical observables for a test particle around a Schwarzschild (or any spherically symmetric) black hole. Specifically, we will particularize the computation for the case of the S2 star in the Galactic center \cite{DeLaurentis2023}. The high ratio between the star's mass ($m\lesssim10M_\odot$) and that of the supermassive black hole Sgr A* ($M\sim4\times10^6M_\odot$) allows to describe the dynamics of the star as a freely falling test particle in the gravitational field of the central object. For this reason, we will assume that S2 moves on a time-like geodesic and will use the \texttt{Orbit} class to obtain its trajectory in PyGRO. The set of classical Keplerian orbital parameters, ($t_P$, $a$, $e$,$i$, $\omega$, $\Omega$), as obtained from astrometric and spectroscopic observations for S2 is reported in the Table 3 of \cite{Gillessen2017}, assuming a mass of Sgr A* of $M=(4.35\pm0.13)\times10^6M_\odot$ and a distance of Earth form the Galactic center of $D=8.33\pm 0.12$ kpc. In our case we can use the values of these Keplerian elements with the same physical meaning that we have provided in Section \ref{sec:orbit}. This results in an integrated four-dimensional trajectory for S2, ($t(\tau)$, $r(\tau)$, $\theta(\tau)$ and $\phi(\tau)$) in the Schwarzschild coordinate system as a function of the rest frame proper time $\tau$ of the star. To make this trajectory comparable with astronomical observations, it is necessary to reconstruct physically observable quantities. The available data for the S-stars in the Galactic center \citep{Gillessen2017} consist of sky-projected astrometric positions for an Earth-based observer and spectroscopically reconstructed line of sight velocities. The information on both observables is thus carried by photons emitted by the star along its orbit and collected by a distant observer. If one wants to approach the problem of reconstructing the astrometric and spectroscopic observables in a fully relativistic way, photon paths represented by null geodesics must be integrated after an appropriate initial-condition-search has been carried out to identify the right null trajectory connecting emitter and observer at each epoch, the so-called ``emitter-observer problem''. The advantage of this approach is that one would be able to take into account at once all the relativistic effects on the star and the photons. This technique, which has been developed in full generality for spherically symmetric space-times in \citet{DellaMonica2023e} and is built upon PyGRO, is quite costly in terms of computational resources. On the other hand, the current observational sensitivity in the Galactic center only allows to detect the leading-order relativistic effects \citep[see e.g. the supplementary materials of][]{Do2019}. For this reason, we can neglect the fully relativistic treatment of light rays and use a semi-classical approach to reconstruct the two observables. A useful way to understand this approach is to consider it as a two-stage process. First, the fully-relativistic timelike geodesic is integrated in a chosen coordinate system (e.g. the usual spherical coordinate) within the curved spacetime. Then, these coordinates are identified with those of a Newtonian spacetime, allowing for the projection of light rays in a flat spacetime using standard Newtonian formulas. This correspons to using the aerial coordinate $r$ as representative of the Euclidean concept of distance between the test particle and the central source, which of course implies attributing a physical meaning to a space-time coordinate that \emph{per se} does not have one. While breaking the general covariance of general relativity, this assumption allows to work with the integrated trajectory as one would do in Newtonian mechanics, by constructing a cartesian frame ($x$, $y$, $z$) of which ($r$, $\theta$, $\phi$) constitute the usual spherical coordinates. From the projection of the star's position and velocity in this cartesian frame one can retrieve the relative positions between Sgr A* and the star projected on the sky-plane (which we consider to be the $xy$ plane) and the kinematic velocity along the line-of-sight, $v_z$. It is fairly easy now to derive the angular separations between Sgr A* and S2, i.e. the experimentally observable astrometric positions,
  \begin{equation}
    \alpha = \frac{y(\tau)}{D},\qquad\delta = \frac{x(\tau)}{D},
    \label{eq:astrometric_observables}
  \end{equation}
  corresponding to right ascension and declination, respectively.

  Positions and velocities of the test particle are computed in terms of proper time. The zero-th component of the integrated space-time trajectory of the particle, however, $t(\tau)$, represents a map between this proper time $\tau$ and the coordinate time $t$, namely (in Schwarzschild coordinates) the time measured by a clock at infinity. This quantity is very useful when reconstructing astronomical observables because we can effectively identify it with the time measured by clocks on Earth, very distant from the gravitational field of Sgr A*, that is used to label the observation epochs. We can thus use the coordinate time as a bookkeeper coordinate to label both the moments of emission of the photons, $t_{\textrm{em}}$, and of arrival at the observer's location, $t_{\textrm{obs}}$. The relation between the two quantities is given by the photon travel time between emission and reception, which is a combination of classical effects and relativistic delays \citep{Damour1986}. For what concerns the S-stars observations, the only effect that gives a non-negligible contribution is the classical Rømer delay. This effects simply accounts for the light-ray propagation time assuming straight-line propagation and can be thus expressed as (we restore here and for the rest of the paragraph the speed of light $c$ for a clearer notation)
  \begin{equation}
    \Delta t_{\textrm{Rømer}} \equiv t_{\textrm{em}}-t_{\textrm{obs}} = \frac{z(t_{\textrm{em}})}{c}.
    \label{eq:roemer}
  \end{equation}
  For inclined orbits, the time-varying character of $z(t_{\textrm{em}})$ induces a periodic modulation on the time of arrival of the photons that one has to take into account. Solving Eq. \eqref{eq:roemer} for $t_{\textrm{obs}}$ gives out a map between observation times and emission times. The zero-th component of the integrated geodesic, $t(\tau)$ relates coordinate times of emission to proper times. Thus, we have an invertible chain of maps that links the proper times with the times of observation and vice versa
  \begin{equation}
    \tau\,\leftrightarrow\, t_{\textrm{em}}(\tau)\,\leftrightarrow\,t_{\textrm{obs}}(t_{\textrm{em}}),
  \end{equation}
  that one can invert numerically to express the proper time as a function of the observation time
  \begin{equation}
    \tau=\tau(t_{\textrm{obs}}).
  \end{equation}
  This allows to reparametrize the reconstructed observables as function of the observation time
  \begin{align}
    \alpha(\tau) &\to \alpha(t_{\textrm{obs}}),\\
    \delta(\tau) &\to \delta(t_{\textrm{obs}}),\\
    v_z(\tau) &\to v_z(t_{\textrm{obs}}),
  \end{align}
  which is a fundamental step for the comparison of synthetic orbits with actual observational data. Due to its high inclination, observations for S2 are drastically affected by Rømer delay. The photons emitted at the apocenter will arrive on Earth almost a week later as compared to a case with no inclination, while at the pericenter photons arrive $\sim 13$ hours earlier.

  \begin{figure}
    \includegraphics[width=\columnwidth]{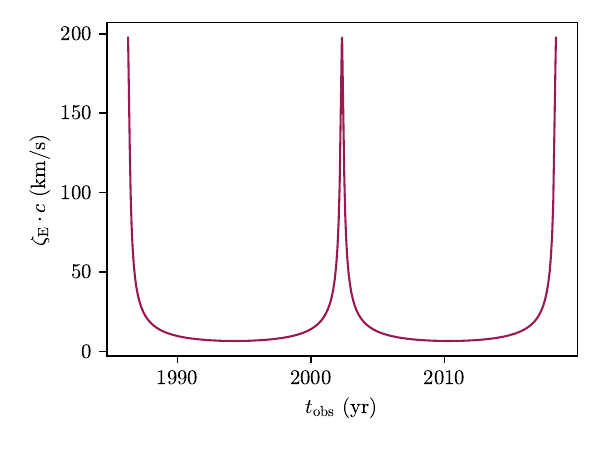}
    \caption{The additional contribution to the apparent line-of-sight velocity given by the combination of special relativistic Doppler effect and gravitational time dilation for the S2 star in the Galactic center. The relativistic redshift component is maximized around pericenter (at $t\sim2002$ and $t\sim2018$), reaching an additional contribution of $\sim200$ km/s \citep{Do2019,GravityCollaboration2018}, where the star goes deeper in the gravitational potential well of Sgr A* and is accelerated to its maximum orbital velocity.}
    \label{fig:gravitational-redshift}
  \end{figure}

  \begin{figure*}
    \includegraphics[width=\textwidth]{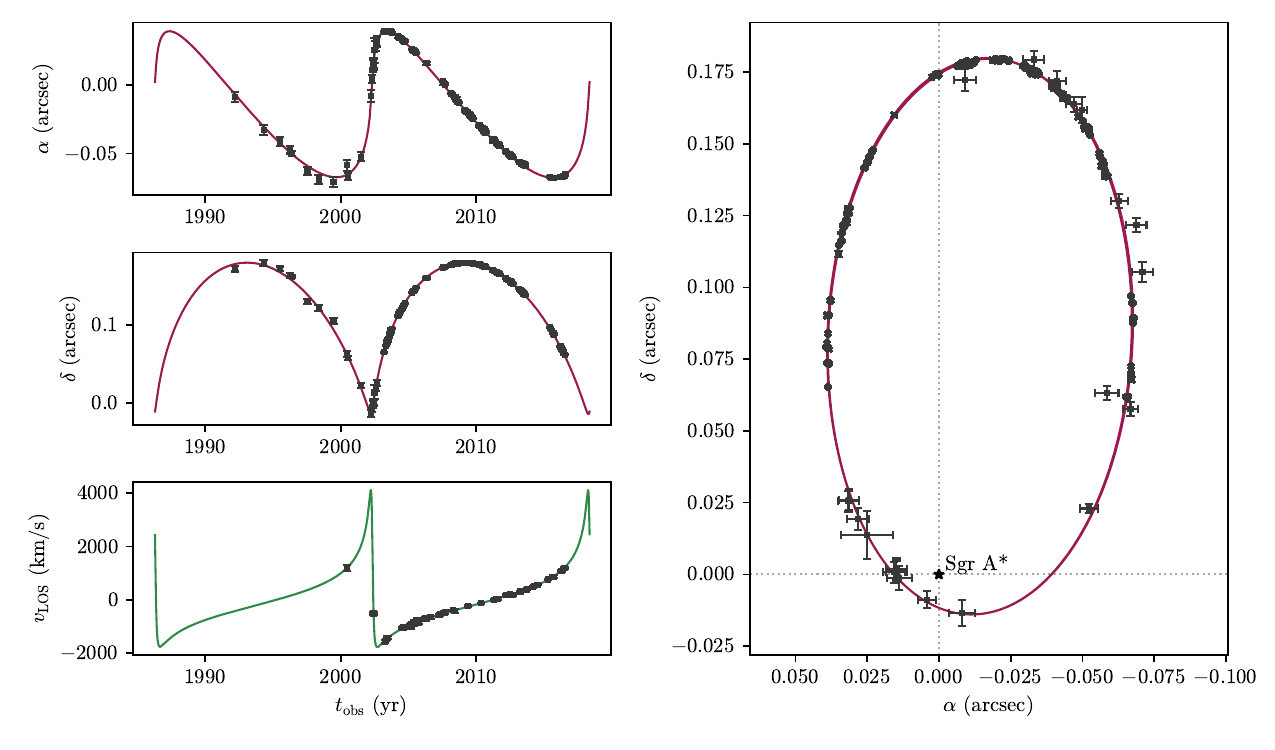}
    \caption{Full set of observables for S2 in the Galactic center, reconstructed from the geodesic integration with PyGRO. In particular, on the left panels we show the right ascension (top panel) and declination (middle panel) relative to Sgr A* of S2 over two full orbital periods, and the fully relativistic line-of-sight velocity (bottom panel). On the right panel we display the full apparent trajectory in the sky relative to Sgr A*. The black dots correspond to the publicly available astrometric and spectroscopic data for S2 from \citet{Gillessen2017} with their corresponding error.}
    \label{fig:S2-observables}
  \end{figure*}

  Let us now move to the line-of-sight velocity. This observable is reconstructed from spectroscopic observations. The apparent redshift $\zeta$ in the star's spectrum is measured and converted into an apparent line-of-sight velocity $v_\textrm{LOS}$, by considering that this shift is produced only due to longitudinal Doppler effect
  \begin{equation}
    \zeta = \frac{\Delta\lambda}{\lambda} - 1 \equiv \frac{v_{\textrm{LOS}}}{c}.
  \end{equation}
  For non-relativistic orbits, the line-of-sight velocity reconstructed with this method matches the kinematic velocity $v_z$ of the celestial body along the line-of-sight. While $v_z$ accounts for most of the apparent redshift,
  \begin{equation}
    \zeta_k(t_{\textrm{obs}}) = \frac{v_z(t_{\textrm{obs}})}{c},
    \label{eq:kinematical-redshift}
  \end{equation}
  the combination of the intense gravitational field produced by Sgr A* and the subsequent high orbital speed brings the S2 star orbit in the relativistic regime. More specifically, the gravitational time dilation due to the proximity to the supermassive source and the transverse Doppler effect (i.e. Lorentz time dilation) due to the high velocity of the star, especially near the pericenter, produce a non-negligible increase in the apparent redshift that must be taken into account to properly reconstruct the spectroscopic observable. The combination of both effects is named Einstein delay. From our fully relativistic integration of the star's motion, we can directly estimate the coordinate-to-proper-time shift, which corresponds to the integrated zero-th component of the 4-velocity
  \begin{equation}
    \zeta_E(t_{\textrm{obs}}) + 1 = \dot{t}(t_{\textrm{obs}}).
    \label{eq:einstein-redshift}
  \end{equation}
  We can then combine this redshift component to the kinematic term in Eq. \eqref{eq:kinematical-redshift}, giving the total apparent redshift
  \begin{equation}
    \frac{v_{\textrm{LOS}}(t_{\textrm{obs}})}{c}+1 = (\zeta_k(t_{\textrm{obs}})+1)(\zeta_E(t_{\textrm{obs}})+1).
    \label{eq:total-redshift}
  \end{equation}

  In Fig. \ref{fig:gravitational-redshift} we report the relativistic redshift reconstructed from the integrated geodesic for the S2 star. The relativistic contribution at the pericenter accounts for an additional $\sim 200$ km/s kick in the apparent line-of-sight velocity that has been successfully measured by high-resolutions spectroscopic observations in the Galactic center \citep{Do2019,GravityCollaboration2018}.

  When the whole procedure explained in this section is taken into account, a fully relativistic set of observables ($\alpha(t_\textrm{obs})$, $\delta(t_\textrm{obs})$, $v_\textrm{LOS}(t_\textrm{obs})$) for a massive test particle around a black hole is obtained. We show in Fig. \ref{fig:S2-observables} the observables reconstructed using PyGRO for the S2 star and compare them to the available sets of astrometric and spectroscopic data for this object \citep{Gillessen2017}. The agreement between the reconstructed observables and the observational data for S2 is striking and showcases the potentiality of PyGRO to link theoretical models for space-time geometries to astrophysically-relevant measurable quantities. This is especially useful to perform model fitting (e.g. by applying Bayesian parameter estimation techniques) of orbital models based on extensions to the Schwarzschild metric, which thus account for modifications to general relativity or the existence of black hole mimickers, leveraging the generality of the orbital integration in PyGRO \citep{DeMartino2021,DellaMonica2022a,DellaMonica2022b,DellaMonica2023a,DellaMonica2023b,DellaMonica2023c,DellaMonica2023d,Cadoni2023,Fernandez2023,DeMoraLosada2025}.

  \subsection{Photon trajectories around a rotating black hole}

  In this section we show how PyGRO can also work outside the family of spherically symmetric space-times. Even though the \texttt{Orbit} class can only work in spherical symmetry, the low level APIs in PyGRO and the \texttt{Observer} class can work in full generality. Here, for example, we study the photon trajectories in the Kerr space-time, whose metric in Boyer-Lindquist coordinates ($t$, $r$, $\theta$, $\phi$) takes the form \citep{Bardeen1972}
  \begin{align}
    ds^2 =& -\left(1-\frac{2Mr}{\Sigma}\right)dt^2-\frac{4Mar\sin^2\theta}{\Sigma}dtd\phi+\frac{\Sigma}{\Delta}dr^2+\Sigma d\theta^2+\nonumber\\
    &+\left(r^2+a^2+\frac{2Ma^2r\sin^2\theta}{\Sigma}\right)\sin^2\theta d\phi^2
    \label{eq:kerr_metric}
  \end{align}
  with $\Sigma \equiv r^2+a^2\cos^2\theta$ and $\Delta \equiv r^2-2Mr+a$.
  Eq. \eqref{eq:kerr_metric} describes the space-time around a rotating black hole of mass $M$ with angular momentum per unit mass $a$. In the limit $M\to0$, the Kerr metric reduces to flat Minkowski space-time in oblate spheroidal coordinates, that are related to standard flat-space cartesian coordiantes via the relations $x=\sqrt{r^2+a^2}\sin\theta\cos\phi$, $y=\sqrt{r^2+a^2}\sin\theta\sin\phi$ and $z=r\cos\theta$. We will use these flat-space limit cartesian coordinates for the purpose of visualization of the geodesics resulting from our calculations.

  For simplicity, we will restrict our consideration to the equatorial plane\footnote{Due to breaking of the spherical symmetry in rotating space-times, not all planes are equal and the equatorial one benefits from additional symmetries, e.g. the fact that geodesic trajectories on the equatorial plane will stick to it for the full duration of the motion.} $\theta=\pi/2$. On this plane, one can uniquely assign initial conditions to null geodesics at a given point by defining an impact parameter $b\equiv L/E$ where $L$ and $E$ are the azimuthal angular momentum and specific energy as defined in Eqs. \eqref{eq:specific-energy}-\eqref{eq:angular-momentum}. A choice of $b$ uniquely fixes\footnote{More precisely, a choice of $b$ fixes the ratio $\dot{\phi}/\dot{t}$ and one can freely assign $\dot{t}$ at the initial time, given the independence of the gedoesic equations on the choice of the affine parameterization.} values of $\dot{t}$ and $\dot{\phi}$ and the conservation of the Lagrangian itself, due to the normalization of null geodesic in Eq. \eqref{eq:geodesic_normalization}, fixes the remaining component $\dot{r}$. We can compute all these quantities symbolically in PyGRO and then invert the relation between $b$ and ($\dot{t}$, $\dot{r}$, $\dot{\phi}$) numerically.This allows to assign initial conditions in PyGRO once an initial position and a value for the impact parameter $b$ have been assigned. An interesting classical result of geodesic motion in the Kerr space-time that we want to recover in PyGRO is the existence of critical impact parameters $b_c$ on the equatorial plane for which the null geodesic results in an unstable photon orbit. It is a known result \citep[see e.g.][]{Chandrasekhar1985} that $b_c$ satisfies the equation
  \begin{equation}
    (b_c+a)^3=27M^2(b_c-a),
    \label{eq:critical_impact_parameter}
  \end{equation}
  for any given value of $a$. For $a=0$ (non-rotating case) the only solution is $b_c=\sqrt{27}M$ which corresponds to the circular photon orbit wit $r=3M$ that we have obtained in Section \ref{sec:photon_orbit} for the Schwarzschild case. For $0<a<1$, due to frame dragging, Eq. \eqref{eq:critical_impact_parameter} has two solutions, $b_{c,\pm}$ with opposite signs, one ($b_{c,+}$) for prograde (null geodesics with $L$ having the same sign of $a$) and the other ($b_{c,-}$) for retrograde (with $L$ and $a$ having opposite signs) null geodesics. This means that there exist two different circular photon orbits on the equatorial plane at radii
  \begin{equation}
    r_\pm = 2M\left[1+\cos\left(\frac{2}{3}\arccos\frac{\pm a}{M}\right)\right].
    \label{eq:circular_photon_orbits_kerr}
  \end{equation}
  Null geodesics with $b_{c,-}<b<b_{c,+}$ will be bound to the central black hole and cross the event horizon. Null geodesics with impact parameter outside this interval will escape to infinity. Null geodesics with $b_{c,\pm}$ will approach asymptotically the unstable circular orbits in Eq. \eqref{eq:circular_photon_orbits_kerr}.

\begin{figure}
    \includegraphics[width=\columnwidth]{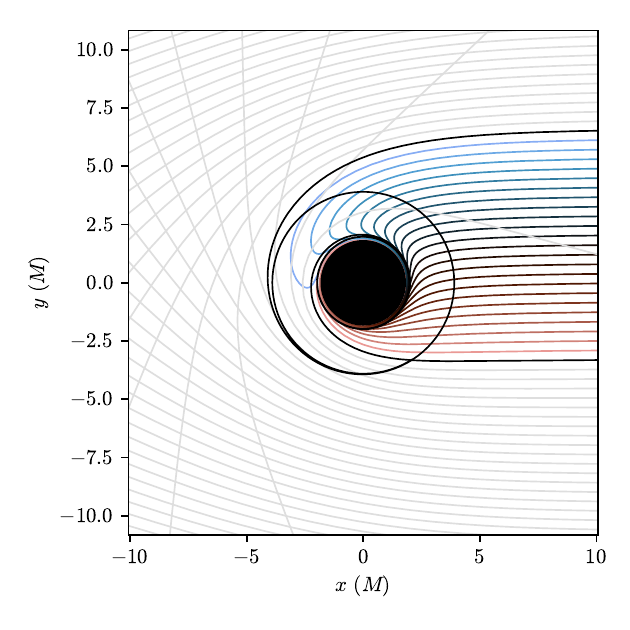}
    \caption{Photon trajectories on the equatorial plane of the Kerr space-time with $a=0.8$. The trajectories are obtained by integrating backward in time from a distant observer on the $x$ axis. The solid black lines correspond to the prograde and retrogrde circular photon orbits, obtained for the critical values, $b_{c,\pm}$, of the impact parameter. Photons with impact parameters between $b_{c,+}$ and $b_{c,-}$ are bound to the black hole and cross the horizon (colored solid lines). Outside this interval photons escape to infinity (gray lines). Due to frame dragging the two critical impact parameters are not symmetrical with respect to the central black hole, which ultimately causes the asymmetry of the Kerr shadow.}
    \label{fig:kerr_photons}
  \end{figure}

  In PyGRO we can reproduce this result by integrating numerically null geodesics with different impact parameters in the space-time in Eq. \eqref{eq:kerr_metric}. In particular, in Fig. \ref{fig:kerr_photons} we show the results for a rapidly rotating black hole with $a=0.8$. We consider and equally spaced set of impact parameters, spanning positive and negative values, and including the two critical impact parameter $b_{c,\pm}$ that we compute by numerically solving Eq. \eqref{eq:critical_impact_parameter}. We fix the starting position of the photons at $r=10^4M$ on the equatorial plane, so to mimic photons received by a distant observer who performs ray-tracing for the Kerr black hole. For each value of $b$ we assign initial conditions as described above and we integrate the geodesic equations backward in time. As expected theoretically, photons with $b_{c,-}<b<b_{c,+}$ (colored lines) cross the horizon, while the others (gray lines) escape to infinity. Geodesics with impact parameter matching the critical values (black lines) approach the unstable circular orbits for the prograde and retrograde cases. The asymmetry of the region corresponding to bound photons with respect to a photon initially aligned with the central black hole ($b=0$) ultimately causes the asymmetry of the Kerr shadow \citep{Cunningham1973, Luminet1979}.

  \subsection{Benchmark of integration accuracy: the radial infall case}

  \begin{figure}
    \includegraphics[width=\columnwidth]{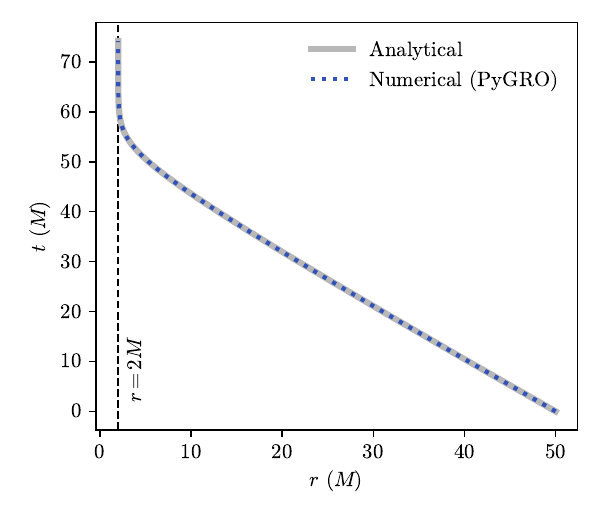}
    \caption{Analytical (gray solid line) and numerical (blue dotted line, obtained with PyGRO) solutions for a radially infalling null geodesic in the Schwarzschild space-time, starting from an initial radial coordinate $r_0=50M$. The two solutions perfectly agree. The dashed vertical line corresponds to the Schwarzschild radius where the coordinate time exhibits a divergence.}
    \label{fig:radial_geodesic}
  \end{figure}

  In this last example, we go back to Schwarzschild metric, Eq. \eqref{eq:schwarzschild_metric}, and consider the case of a radially infalling null geodesic, for which an analytical solution is known. We compare the results obtained with PyGRO with those from the analytical formula and check the accuracy of the integration of the geodesic equation within PyGRO. Also, we perform some convergence check based on the conserved norm of the tangent four-vector to the integrated geodesic.

  Radial null geodesics are the simplest case of trajectories that one can consider in the space-time described by Eq. \eqref{eq:schwarzschild_metric}. In this case, due to the spherical symmetry of the problem, the geodesic equations reduce only to the temporal and radial components, enabling a complete analytical solution to be obtained, which has the form \citep{Chandrasekhar1985}
  \begin{equation}
    t(r) = \mp (r+2M\log(r-2M)) \pm (r_0+2M\log(r_0-2M))
    \label{eq:radial_geodesic_analytical}
  \end{equation}
  where $r_0$ is the initial radial coordinate and the solution corresponding to the first sign being $-$ ($+$) corresponds to an inward (outward) pointing geodesic. The appearance of a logarithm exhibits one of the most peculiar behaviors of the dynamics of particles in the near-horizon zone. The coordinate time of an infalling geodesic tends to diverge as it approaches the horizon at $r=2M$. This divergence means that a very distant physical observer will never see the infalling particle cross the horizon. We are going to reproduce the same configuration in PyGRO. To do so we consider an \texttt{Observer} defined as in Eqs. \eqref{eq:observer-frame-1}-\eqref{eq:observer-frame-4}, at a radial coordinate $r_0$ on the equatorial plane of the Schwarzschild space-time. We then consider a null geodesic fired at the observer's location along the $f_1$ axis, so that it points radially inward. We carry on the integration up to a given stopping criterion (e.g. in this case we stop the integration when the geodesic reaches radial coordinate which is right above the event horizon radius $r=2M$). We than apply the formula in Eq. \eqref{eq:radial_geodesic_analytical} to obtain the analytical solution and we compare the two results at the same radii.

  In Fig. \ref{fig:radial_geodesic} we show the $t(r)$ profiles obtained in the two cases for an initial radius of $r_0 = 50M$. The two curves perfectly agree, with the numerical result obtained in PyGRO reproducing the divergence of the coordinate time at $r=2M$. To better quantify the agreement between the two estimations, we perform a convergence check, whose results we report in Fig. \ref{fig:circular_orbits}. Our analysis shows that: (i) by reducing the integration tolerance the discrepancy between the result of the numerical integration and the analytical formula in Eq. \eqref{fig:radial_geodesic} converges to zero and always matches, on average, the order of magnitude of the integration tolerance; (ii) the amount of constraint violation (we focus specifically on the normalization of the tangent four-vector to the geodesic reported in Eq. \eqref{eq:geodesic_normalization}) tends to zero as well, as the integration tolerance is improved. These results show not only that the numerical routine implemented in PyGRO allows for accurate integration of the geodesic equations, but also that the amount of error and constraint violation in the integration can be controlled effectively by setting appropriate precision and accuracy goals in the integrator. Additionally, we have checked that the best results, in terms of integration accuracy and computational time, are obtained employing a Dormand-Prince5(4) integrator, which is thus set as the default choice in PyGRO.

  \begin{figure*}
    \includegraphics[width=\textwidth]{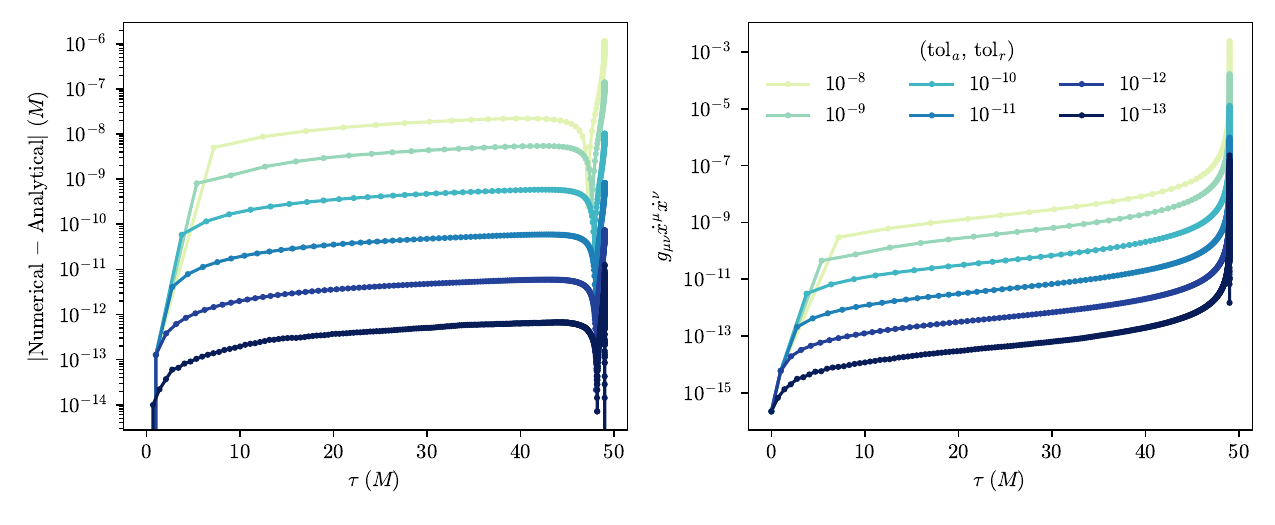}
    \caption{Convergence check of the numerical integration routine in PyGRO. In particular, for the same case illustrated in Fig. \ref{fig:radial_geodesic}, with $r_0=50M$, we compute the difference (left panel) between the known analytical results, Eq. \eqref{eq:radial_geodesic_analytical} and the results obtained in PyGRO, for different values of the precision and accuracy tolerances (Eq. \eqref{eq:tolerances}) that we vary over six orders of magnitude. The plots shows the convergence of the numerical result to the analytical one as the integration tolerances are reduced. Moreover, we compute on the integrated geodesic the norm of the tangent four-vector (right panel) which, following Eq. \eqref{eq:geodesic_normalization}, should be identically zero. The plot, thus reports the amount of constraint violation which shows convergence to zero as the integration precision is improved, thus validating the numerical routines implemented in PyGRO.}
    \label{fig:convergence}
  \end{figure*}

  \subsection{Integration performances}
  \label{sec:performances}

  In this section we report the performances of PyGRO, showcasing the runtimes for all the examples described in the previous sections. PyGRO is light-weight and well optimized to run on any machine. All the examples and tutorials on the official documentation can be run on a normal laptop or even on mobile platforms (e.g. Juno for the iOS ecosystem or Pydroid for Android platforms). All the runtimes reported here have been obtained on a M1 MacBook Pro with 8GBs of RAM.

  The symbolic differentiation and matrix operations required for the initialization of the \texttt{Metric} object, upon which all the symbolic manipulations are done to compute the Christoffel symbols and the geodesic equations (see Section \ref{sec:metric}), while introducing an overhead compared to other codes \citep[e.g.][]{Vincent2011} where all derived quantities are hardcoded, is usually fast, with the operation of computing the helper functions to normalize the initial tengent four-vector being the most computationally expensive. In general, initializing a \texttt{Metric} object takes about $\sim\mathcal{O}(1\textrm{s})$. The initialization of a \texttt{GeodesicEngine} can require slightly longer, normally $\sim\mathcal{O}(10\textrm{s})$, due to the pre-compilation of the C callables for the geodesic equations. This overhead, represents the trade-off between flexibility and performance that allows PyGRO to work with arbitrary space-time geometries without requiring manual derivations. If the \emph{lambdify} wrapper is used instead (see Section \ref{sec:geodesic_engine}), the overhead is much shorter, usually $\sim\mathcal{O}(0.1\textrm{s})$, but will bring to a loss in performances on integration.

  Integrating a \texttt{Geodesic} in PyGRO can be really fast depending on the specific configuration of the trajectory (e.g. how close it gets to an horizon, where a smaller step size is required to attain a given precision tolerance). From all the examples that we have shown throughout Section \ref{sec:examples} for the Schwarzschild metric, we have focused on seven specific cases for which we systematically assess the integration performances of PyGRO. The cases taken into considerations are:
  \begin{enumerate}
    \item[(1)] The precessing time-like orbit in the Schwarzschild space-time shown in Fig. \ref{fig:orbital-precession-1} with $e = 0.8$ and $a=200M$, integrated over 10 full revolutions.
    \item[(2)] The unstable circular orbit for a time-like geodesic with $a=5M$ shown in the left panel of Fig. \ref{fig:circular_orbits}. As in the last case, we carry on the integration for 10 full revolutions.
    \item[(3)] A stable circular orbit for a time-like geodesic with $a=10M$ shown in the left panel of Fig. \ref{fig:circular_orbits}.  Again, we consider 10 full revolutions.
    \item[(4)] One of the plunging null geodesics shown in Fig. \ref{fig:photon_orbits}. In particular, we consider the geodesic with $r_0=2.1M$ and we use as stopping criterion the condition $r > 2.001M$ (to stop the integration when the geodesic is right above the horizon).
    \item[(5)] The unstable photon orbit with $r_0=3M$ shown in Fig. \ref{fig:photon_orbits}. In this case, we stop the integration after 10 full revolutions.
    \item[(6)] One of escaping photon trajectories from Fig. \ref{fig:photon_orbits}. We focus in particular on the geodesic starting at $r_0=6M$. The integration is stopped at $s=1000M$, being $s$ the affine parameter on the null geodesic.
    \item[(7)] The radially infalling geodesic in Fig. \ref{fig:radial_geodesic} starting at $r_0=50M$. As in case (4) we consider as condition for the stopping criterion $r > 2.001M$.
  \end{enumerate}

  In Table \ref{tab:runtimes} we report the runtimes obtained for all the cases described above, for three different values of the integration tolerance in Eq. \eqref{eq:tolerances}, using the DormandPrince5(4) integrator. The values reported in the table correspond to the mean runtime over $10^2$ repetitions of the geodesic integration, with the error computed as their standard deviation.

  \begin{table}
    \setlength{\tabcolsep}{23pt}
    \footnotesize
    \renewcommand{\arraystretch}{1.4}
    \caption{Integration performances for all the examples in Section \ref{sec:examples}.}
    \begin{tabular}{c|cc}
      \hline
      Example & Tolerance & Runtime (ms) \\ \hline
      \multirow{3}{*}{(1)} & $10^{-8}$ & $77.5\pm3.6$\\
      & $10^{-11}$ &  $202 \pm 22$ \\
      & $10^{-13}$ &  $405\pm28$ \\ \hline
      \multirow{3}{*}{(2)} & $10^{-8}$ & $2.41\pm0.23$\\
      & $10^{-11}$ &  $5.28\pm0.49$ \\
      & $10^{-13}$ &  $9.3\pm2.5$ \\ \hline
      \multirow{3}{*}{(3)} & $10^{-8}$ & $1.97\pm0.04$\\
      & $10^{-11}$ &  $2.96\pm0.17$ \\
      & $10^{-13}$ &  $4.19\pm0.08$ \\ \hline
      \multirow{3}{*}{(4)} & $10^{-8}$ & $8.6\pm0.5$ \\
      & $10^{-11}$ &  $31.1\pm1.9$ \\
      & $10^{-13}$ &  $74.8\pm0.5$ \\ \hline
      \multirow{3}{*}{(5)} & $10^{-8}$ & $17.5\pm0.47$ \\
      & $10^{-11}$ &  $39.1\pm3.1$ \\
      & $10^{-13}$ &  $97.8\pm1.6$ \\ \hline
      \multirow{3}{*}{(6)} & $10^{-8}$ & $7.75\pm0.40$\\
      & $10^{-11}$ &  $26.3\pm3.2$ \\
      & $10^{-13}$ &  $61.4\pm1.6$ \\ \hline
      \multirow{3}{*}{(7)} & $10^{-8}$ & $18.5\pm0.7$\\
      & $10^{-11}$ &  $50.5\pm4.1$ \\
      & $10^{-13}$ &  $110.0\pm3.4$ \\ \hline
    \end{tabular}
    \tablefoot{The examples considered correspond to: (1) Precessing orbit shown in Fig. \ref{fig:orbital-precession-1}; (2) Unstable circular orbit in Fig. \ref{fig:circular_orbits} (left panel); (3) Stable circular orbit in Fig. \ref{fig:circular_orbits} (right panel); (4) Plunging null geodesic with $r_0=2.1M$ in Fig. \ref{fig:photon_orbits}; (5) Unstable photon orbit with $r_0=3M$ in Fig. \ref{fig:photon_orbits}; (6) Escaping null geodesic with $r_0=6M$ in Fig. \ref{fig:photon_orbits}; (7) radially infalling geodesic in Fig. \ref{fig:radial_geodesic}.}
    \label{tab:runtimes}
  \end{table}

  \section{Discussion and conclusions}

  The study of relativistic orbits plays a central role in our understanding of the nature of gravity and our ability to test the predictions of general relativity. Over the last century, the equivalence principle and classical tests of gravity have evolved from basic confirmations of Einstein's theory to precision measurements in extreme astrophysical environments \citep{Will2014}. Recent advances in experimental gravitation -- such as the detailed monitoring of the S-stars at the Galactic center \citep{DeLaurentis2023}, the groundbreaking imaging of black holes by the Event Horizon Telescope \citep{EHT2019, EHT2022}, and the potential discovery of pulsars orbiting a supermassive black hole \citep{DellaMonica2025} -- have ushered in a new era of strong-field gravitational physics.

  These advancements call for modern computational tools that can bridge theoretical predictions with observations in increasingly complex and general settings. PyGRO was developed with this goal in mind: to provide an open-source, fast, and highly customizable framework for the numerical integration of geodesic equations in any analytic four-dimensional space-time. By combining symbolic and numerical methods, PyGRO allows users to explore relativistic dynamics in a wide range of scenarios, from classical tests in weak gravitational fields to orbits around compact objects in the strong-field regime. The examples presented in this article and the in the official documentation demonstrate the robustness of PyGRO, validating its methodology by successfully reproducing key results of general relativity, while extensive use of the code in past scientific studies \citep{DeMartino2021,DellaMonica2022a,DellaMonica2022b,DellaMonica2023a,DellaMonica2023b,DellaMonica2023c,DellaMonica2023d,Cadoni2023,Fernandez2023,DeMoraLosada2025} demonstrates its great flexibility. Beyond these benchmarks, the modular and open-source nature of PyGRO enables the future implementation in the code of extension and extra features, to further improve its range of applications. For example, we plan in the near future to extend the capabilities of integrating general relativistic orbits with a parameterization based on the Keplerian orbital elements of classical celestial mechanics, also to the case of axisymmetric space-times, namely rotating black holes. While we have shown in this article that the low-level APIs in PyGRO offer sufficient generality to be easily applied to axisymmetric space-times, like the Kerr solution, the mapping between Keplerian elements and initial conditions in this case requires further theoretical modelling, especially for orbits outside the equatorial plane. Additionally, the implementation of polarized general relativistic ray-tracing techniques \citep[like the ones presented in][]{Aimar2024} could significantly extend the scope of PyGRO, enabling tests of the properties of relativistic astrophysical plasmas.

  PyGRO opens new possibilities for the broader astrophysics community, offering a powerful tool for both theoretical studies and the analysis of observational data, in preparation for next-generation experiments \citep{GravityPlus2022,Keane2015}.

  \begin{acknowledgements}
    PyGRO is the culmination of the work carried out during my PhD studies at the University of Salamanca. I am sincerely indebted to my supervisor Ivan de Martino for his guidance. I also thank Frédéric Vincent for providing valuable insights and suggesting relevant improvements on the final manuscript. Moreover, I thank Gabriel Sánchez Pérez and Diego Martín González for testing my code and suggesting some relevant implementations. Finally, I acknowledge financial support from the  grant PID2021-122938NB-I00 funded by MCIN/AEI/10.13039/501100011033 and by ``ERDF A way of making Europe'', from Consejeria de Educación de la Junta de Castilla y León and the European Social Fund +.
  \end{acknowledgements}

  \bibliographystyle{aa}
  \bibliography{biblio}

\begin{thebibliography}{59}
\expandafter\ifx\csname natexlab\endcsname\relax\def\natexlab#1{#1}\fi

\bibitem[{{Abbott} {et~al.}(2016){Abbott}, {Abbott}, {Abbott}, {Abernathy}, {Acernese}, {Ackley}, {Adams}, {Adams}, {Addesso}, {Adhikari}, {Adya}, {Affeldt}, {Agathos}, {Agatsuma}, {Aggarwal}, {Aguiar}, {Aiello}, {Ain}, {Ajith}, {Allen}, {Allocca}, {Altin}, {Anderson}, {Anderson}, {Arai}, {Arain}, {Araya}, {Arceneaux}, {Areeda}, {Arnaud}, {Arun}, {Ascenzi}, {Ashton}, {Ast}, {Aston}, {Astone}, {Aufmuth}, {Aulbert}, {Babak}, {Bacon}, {Bader}, {Baker}, {Baldaccini}, {Ballardin}, {Ballmer}, {Barayoga}, {Barclay}, {Barish}, {Barker}, {Barone}, {Barr}, {Barsotti}, {Barsuglia}, {Barta}, {Bartlett}, {Barton}, {Bartos}, {Bassiri}, {Basti}, {Batch}, {Baune}, {Bavigadda}, {Bazzan}, {Behnke}, {Bejger}, {Belczynski}, {Bell}, {Bell}, {Berger}, {Bergman}, {Bergmann}, {Berry}, {Bersanetti}, {Bertolini}, {Betzwieser}, {Bhagwat}, {Bhandare}, {Bilenko}, {Billingsley}, {Birch}, {Birney}, {Birnholtz}, {Biscans}, {Bisht}, {Bitossi}, {Biwer}, {Bizouard}, {Blackburn}, {Blair}, {Blair}, {Blair}, {Bloemen}, {Bock}, {Bodiya}, {Boer}, {Bogaert}, {Bogan}, {Bohe}, {Bojtos}, {Bond}, {Bondu}, {Bonnand}, {Boom}, {Bork}, {Boschi}, {Bose}, {Bouffanais}, {Bozzi}, {Bradaschia}, {Brady}, {Braginsky}, {Branchesi}, {Brau}, {Briant}, {Brillet}, {Brinkmann}, {Brisson}, {Brockill}, {Brooks}, {Brown}, {Brown}, {Brown}, {Buchanan}, {Buikema}, {Bulik}, {Bulten}, {Buonanno}, {Buskulic}, {Buy}, {Byer}, {Cabero}, {Cadonati}, {Cagnoli}, {Cahillane}, {Bustillo}, {Callister}, {Calloni}, {Camp}, {Cannon}, {Cao}, {Capano}, {Capocasa}, {Carbognani}, {Caride}, {Diaz}, {Casentini}, {Caudill}, {Cavagli{\`a}}, {Cavalier}, {Cavalieri}, {Cella}, {Cepeda}, {Baiardi}, {Cerretani}, {Cesarini}, {Chakraborty}, {Chalermsongsak}, {Chamberlin}, {Chan}, {Chao}, {Charlton}, {Chassande-Mottin}, {Chen}, {Chen}, {Cheng}, {Chincarini}, {Chiummo}, {Cho}, {Cho}, {Chow}, {Christensen}, {Chu}, {Chua}, {Chung}, {Ciani}, {Clara}, {Clark}, {Cleva}, {Coccia}, {Cohadon}, {Colla}, {Collette}, {Cominsky}, {Constancio}, {Conte}, {Conti}, {Cook}, {Corbitt}, {Cornish}, {Corsi}, {Cortese}, {Costa}, {Coughlin}, {Coughlin}, {Coulon}, {Countryman}, {Couvares}, {Cowan}, {Coward}, \& {Cowart}}]{Abbott2016}
{Abbott}, B.~P., {Abbott}, R., {Abbott}, T.~D., {et~al.} 2016, \prl, 116, 061102

\bibitem[{{Aimar} {et~al.}(2024){Aimar}, {Paumard}, {Vincent}, {Gourgoulhon}, \& {Perrin}}]{Aimar2024}
{Aimar}, N., {Paumard}, T., {Vincent}, F.~H., {Gourgoulhon}, E., \& {Perrin}, G. 2024, Classical and Quantum Gravity, 41, 095010

\bibitem[{{Bardeen} {et~al.}(1972){Bardeen}, {Press}, \& {Teukolsky}}]{Bardeen1972}
{Bardeen}, J.~M., {Press}, W.~H., \& {Teukolsky}, S.~A. 1972, \apj, 178, 347

\bibitem[{{Bernstein} {et~al.}(1993){Bernstein}, {Tyson}, \& {Kochanek}}]{Bernstein1993}
{Bernstein}, G.~M., {Tyson}, J.~A., \& {Kochanek}, C.~S. 1993, \aj, 105, 816

\bibitem[{{Bolton}(1972)}]{Bolton1972}
{Bolton}, C.~T. 1972, \nat, 235, 271

\bibitem[{{Cadoni} {et~al.}(2023){Cadoni}, {De Laurentis}, {De Martino}, {Della Monica}, {Oi}, \& {Sanna}}]{Cadoni2023}
{Cadoni}, M., {De Laurentis}, M., {De Martino}, I., {et~al.} 2023, \prd, 107, 044038

\bibitem[{Chandrasekhar(1985)}]{Chandrasekhar1985}
Chandrasekhar, S. 1985, {The mathematical theory of black holes}

\bibitem[{{Cunningham} \& {Bardeen}(1973)}]{Cunningham1973}
{Cunningham}, C.~T. \& {Bardeen}, J.~M. 1973, \apj, 183, 237

\bibitem[{{Damour} \& {Deruelle}(1985)}]{Damour1985}
{Damour}, T. \& {Deruelle}, N. 1985, Annales de L'Institut Henri Poincare Section (A) Physique Theorique, 43, 107

\bibitem[{{Damour} \& {Deruelle}(1986)}]{Damour1986}
{Damour}, T. \& {Deruelle}, N. 1986, Annales de L'Institut Henri Poincare Section (A) Physique Theorique, 44, 263

\bibitem[{{De Laurentis} {et~al.}(2023){De Laurentis}, {de Martino}, \& {Della Monica}}]{DeLaurentis2023}
{De Laurentis}, M., {de Martino}, I., \& {Della Monica}, R. 2023, Reports on Progress in Physics, 86, 104901

\bibitem[{{De Martino} {et~al.}(2021){De Martino}, {della Monica}, \& {De Laurentis}}]{DeMartino2021}
{De Martino}, I., {della Monica}, R., \& {De Laurentis}, M. 2021, \prd, 104, L101502

\bibitem[{{de Mora Losada} {et~al.}(2025){de Mora Losada}, {Della Monica}, {de Martino}, \& {De Laurentis}}]{DeMoraLosada2025}
{de Mora Losada}, V., {Della Monica}, R., {de Martino}, I., \& {De Laurentis}, M. 2025, \aap, 694, A280

\bibitem[{{Della Monica} \& {de Martino}(2022)}]{DellaMonica2022a}
{Della Monica}, R. \& {de Martino}, I. 2022, \jcap, 2022, 007

\bibitem[{{Della Monica} \& {de Martino}(2023{\natexlab{a}})}]{DellaMonica2023d}
{Della Monica}, R. \& {de Martino}, I. 2023{\natexlab{a}}, \prd, 108, L101303

\bibitem[{{Della Monica} \& {de Martino}(2023{\natexlab{b}})}]{DellaMonica2023a}
{Della Monica}, R. \& {de Martino}, I. 2023{\natexlab{b}}, \aap, 670, L4

\bibitem[{{Della Monica} \& {de Martino}(2025)}]{DellaMonica2025}
{Della Monica}, R. \& {de Martino}, I. 2025, to be submitted to Phys. Rev. D, arXiv:2501.03912

\bibitem[{{Della Monica} {et~al.}(2022){Della Monica}, {de Martino}, \& {de Laurentis}}]{DellaMonica2022b}
{Della Monica}, R., {de Martino}, I., \& {de Laurentis}, M. 2022, \mnras, 510, 4757

\bibitem[{{Della Monica} {et~al.}(2023{\natexlab{a}}){Della Monica}, {De Martino}, \& {De Laurentis}}]{DellaMonica2023e}
{Della Monica}, R., {De Martino}, I., \& {De Laurentis}, M. 2023{\natexlab{a}}, \mnras, 524, 3782

\bibitem[{{Della Monica} {et~al.}(2023{\natexlab{b}}){Della Monica}, {De Martino}, \& {De Laurentis}}]{DellaMonica2023c}
{Della Monica}, R., {De Martino}, I., \& {De Laurentis}, M. 2023{\natexlab{b}}, \mnras, 524, 3782

\bibitem[{{Della Monica} {et~al.}(2023{\natexlab{c}}){Della Monica}, {de Martino}, {Vernieri}, \& {de Laurentis}}]{DellaMonica2023b}
{Della Monica}, R., {de Martino}, I., {Vernieri}, D., \& {de Laurentis}, M. 2023{\natexlab{c}}, \mnras, 519, 1981

\bibitem[{{Di Francesco} {et~al.}(2019){Di Francesco}, {Chalmers}, {Denman}, {Fissel}, {Friesen}, {Gaensler}, {Hlavacek-Larrondo}, {Kirk}, {Matthews}, {O'Dea}, {Robishaw}, {Rosolowsky}, {Rupen}, {Sadavoy}, {Sa-Harb}, {Sivakoff}, {Tahani}, {van der Marel}, {White}, \& {Wilson}}]{DiFrancesco2019}
{Di Francesco}, J., {Chalmers}, D., {Denman}, N., {et~al.} 2019, in Canadian Long Range Plan for Astronomy and Astrophysics White Papers, Vol. 2020, 32

\bibitem[{Dicke(1964)}]{Dicke1964}
Dicke, R.~H. 1964, in {Les Houches Summer Shcool of Theoretical Physics}: {Relativity, Groups and Topology}, 165--316

\bibitem[{{Do} {et~al.}(2019){Do}, {Hees}, {Ghez}, {Martinez}, {Chu}, {Jia}, {Sakai}, {Lu}, {Gautam}, {O'Neil}, {Becklin}, {Morris}, {Matthews}, {Nishiyama}, {Campbell}, {Chappell}, {Chen}, {Ciurlo}, {Dehghanfar}, {Gallego-Cano}, {Kerzendorf}, {Lyke}, {Naoz}, {Saida}, {Sch{\"o}del}, {Takahashi}, {Takamori}, {Witzel}, \& {Wizinowich}}]{Do2019}
{Do}, T., {Hees}, A., {Ghez}, A., {et~al.} 2019, Science, 365, 664

\bibitem[{Dyson {et~al.}(1920)Dyson, Eddington, \& Davidson}]{Eddington1920}
Dyson, F.~W., Eddington, A.~S., \& Davidson, C. 1920, Philosophical Transactions of the Royal Society of London. Series A, Containing Papers of a Mathematical or Physical Character, 220, 291

\bibitem[{{Eckart} \& {Genzel}(1996)}]{Eckart1996}
{Eckart}, A. \& {Genzel}, R. 1996, \nat, 383, 415

\bibitem[{{Einstein}(1915)}]{Einstein1915a}
{Einstein}, A. 1915, Sitzungsberichte der K{\"o}niglich Preu{\ss}ischen Akademie der Wissenschaften (Berlin), 844

\bibitem[{Einstein(1915)}]{Einstein1915b}
Einstein, A. 1915, Sitzungsber. Preuss. Akad. Wiss. Berlin (Math. Phys. ), 1915, 831

\bibitem[{{Event Horizon Telescope Collaboration} {et~al.}(2024){Event Horizon Telescope Collaboration}, {Akiyama}, {Alberdi}, {Alef}, {Algaba}, {Anantua}, {Asada}, {Azulay}, {Bach}, {Baczko}, {Ball}, {Balokovic}, {Bandyopadhyay}, {Barrett}, {Baub{\"o}ck}, {Benson}, {Bintley}, {Blackburn}, {Blundell}, {Bouman}, {Bower}, {Boyce}, {Bremer}, {Brinkerink}, {Brissenden}, {Britzen}, {Broderick}, {Broguiere}, {Bronzwaer}, {Bustamante}, {Byun}, {Carlstrom}, {Ceccobello}, {Chael}, {Chan}, {Chang}, {Chatterjee}, {Chatterjee}, {Chen}, {Chen}, {Cheng}, {Cho}, {Christian}, {Conroy}, {Conway}, {Cordes}, {Crawford}, {Crew}, {Cruz-Osorio}, {Cui}, {Dahale}, {Davelaar}, {De Laurentis}, {Deane}, {Dempsey}, {Desvignes}, {Dexter}, {Dhruv}, {Dihingia}, {Doeleman}, {Dougal}, {Dzib}, {Eatough}, {Emami}, {Falcke}, {Farah}, {Fish}, {Fomalont}, {Ford}, {Foschi}, {Fraga-Encinas}, {Freeman}, {Friberg}, {Fromm}, {Fuentes}, {Galison}, {Gammie}, {Garc{\'\i}a}, {Gentaz}, {Georgiev}, {Goddi}, {Gold}, {G{\'o}mez-Ruiz}, {G{\'o}mez}, {Gu}, {Gurwell}, {Hada}, {Haggard}, {Haworth}, {Hecht}, {Hesper}, {Heumann}, {Ho}, {Ho}, {Honma}, {Huang}, {Huang}, {Hughes}, {Ikeda}, {Impellizzeri}, {Inoue}, {Issaoun}, {James}, {Jannuzi}, {Janssen}, {Jeter}, {Jiang}, {Jim{\'e}nez-Rosales}, {Johnson}, {Jorstad}, {Joshi}, {Jung}, {Karami}, {Karuppusamy}, {Kawashima}, {Keating}, {Kettenis}, {Kim}, {Kim}, {Kim}, {Kim}, {Kino}, {Koay}, {Kocherlakota}, {Kofuji}, {Koch}, {Koyama}, {Kramer}, {Kramer}, {Kramer}, {Krichbaum}, {Kuo}, {La Bella}, {Lauer}, {Lee}, {Lee}, {Leung}, {Levis}, {Li}, {Lico}, {Lindahl}, {Lindqvist}, {Lisakov}, {Liu}, {Liu}, {Liuzzo}, {Lo}, {Lobanov}, {Loinard}, {Lonsdale}, {Lowitz}, {Lu}, {MacDonald}, {Mao}, {Marchili}, {Markoff}, {Marrone}, {Marscher}, {Mart{\'\i}-Vidal}, {Matsushita}, {Matthews}, {Medeiros}, {Menten}, {Michalik}, {Mizuno}, {Mizuno}, {Moran}, {Moriyama}, {Moscibrodzka}, {Mulaudzi}, {M{\"u}ller}, {M{\"u}ller}, {Mus}, {Musoke}, {Myserlis}, {Nadolski}, {Nagai}, {Nagar}, {Nakamura}, {Narayanan}, {Natarajan}, {Nathanail}, {Fuentes}, {Neilsen}, {Neri}, {Ni}, {Noutsos}, {Nowak}, {Oh}, {Okino}, {Olivares}, {Ortiz-Le{\'o}n}, {Oyama}, {{\"O}zel}, {Palumbo}, {Paraschos}, {Park}, {Parsons}, {Patel}, \& {Pen}}]{EventHorizonTelescope2024}
{Event Horizon Telescope Collaboration}, {Akiyama}, K., {Alberdi}, A., {et~al.} 2024, \apjl, 964, L25

\bibitem[{{Event Horizon Telescope Collaboration} {et~al.}(2022){Event Horizon Telescope Collaboration}, {Akiyama}, {Alberdi}, {Alef}, {Algaba}, {Anantua}, {Asada}, {Azulay}, {Bach}, {Baczko}, {Ball}, {Balokovi{\'c}}, {Barrett}, {Baub{\"o}ck}, {Benson}, {Bintley}, {Blackburn}, {Blundell}, {Bouman}, {Bower}, {Boyce}, {Bremer}, {Brinkerink}, {Brissenden}, {Britzen}, {Broderick}, {Broguiere}, {Bronzwaer}, {Bustamante}, {Byun}, {Carlstrom}, {Ceccobello}, {Chael}, {Chan}, {Chatterjee}, {Chatterjee}, {Chen}, {Chen}, {Cheng}, {Cho}, {Christian}, {Conroy}, {Conway}, {Cordes}, {Crawford}, {Crew}, {Cruz-Osorio}, {Cui}, {Davelaar}, {De Laurentis}, {Deane}, {Dempsey}, {Desvignes}, {Dexter}, {Dhruv}, {Doeleman}, {Dougal}, {Dzib}, {Eatough}, {Emami}, {Falcke}, {Farah}, {Fish}, {Fomalont}, {Ford}, {Fraga-Encinas}, {Freeman}, {Friberg}, {Fromm}, {Fuentes}, {Galison}, {Gammie}, {Garc{\'\i}a}, {Gentaz}, {Georgiev}, {Goddi}, {Gold}, {G{\'o}mez-Ruiz}, {G{\'o}mez}, {Gu}, {Gurwell}, {Hada}, {Haggard}, {Haworth}, {Hecht}, {Hesper}, {Heumann}, {Ho}, {Ho}, {Honma}, {Huang}, {Huang}, {Hughes}, {Ikeda}, {Impellizzeri}, {Inoue}, {Issaoun}, {James}, {Jannuzi}, {Janssen}, {Jeter}, {Jiang}, {Jim{\'e}nez-Rosales}, {Johnson}, {Jorstad}, {Joshi}, {Jung}, {Karami}, {Karuppusamy}, {Kawashima}, {Keating}, {Kettenis}, {Kim}, {Kim}, {Kim}, {Kim}, {Kino}, {Koay}, {Kocherlakota}, {Kofuji}, {Koch}, {Koyama}, {Kramer}, {Kramer}, {Krichbaum}, {Kuo}, {La Bella}, {Lauer}, {Lee}, {Lee}, {Leung}, {Levis}, {Li}, {Lico}, {Lindahl}, {Lindqvist}, {Lisakov}, {Liu}, {Liu}, {Liuzzo}, {Lo}, {Lobanov}, {Loinard}, {Lonsdale}, {Lu}, {Mao}, {Marchili}, {Markoff}, {Marrone}, {Marscher}, {Mart{\'\i}-Vidal}, {Matsushita}, {Matthews}, {Medeiros}, {Menten}, {Michalik}, {Mizuno}, {Mizuno}, {Moran}, {Moriyama}, {Moscibrodzka}, {M{\"u}ller}, {Mus}, {Musoke}, {Myserlis}, {Nadolski}, {Nagai}, {Nagar}, {Nakamura}, {Narayan}, {Narayanan}, {Natarajan}, {Nathanail}, {Fuentes}, {Neilsen}, {Neri}, {Ni}, {Noutsos}, {Nowak}, {Oh}, {Okino}, {Olivares}, {Ortiz-Le{\'o}n}, {Oyama}, {{\"O}zel}, {Palumbo}, {Paraschos}, {Park}, {Parsons}, {Patel}, {Pen}, {Pesce}, {Pi{\'e}tu}, {Plambeck}, {PopStefanija}, {Porth}, {P{\"o}tzl}, {Prather}, {Preciado-L{\'o}pez}, \& {Psaltis}}]{EHT2022}
{Event Horizon Telescope Collaboration}, {Akiyama}, K., {Alberdi}, A., {et~al.} 2022, \apjl, 930, L12

\bibitem[{{Event Horizon Telescope Collaboration} {et~al.}(2019){Event Horizon Telescope Collaboration}, {Akiyama}, {Alberdi}, {Alef}, {Asada}, {Azulay}, {Baczko}, {Ball}, {Balokovi{\'c}}, {Barrett}, {Bintley}, {Blackburn}, {Boland}, {Bouman}, {Bower}, {Bremer}, {Brinkerink}, {Brissenden}, {Britzen}, {Broderick}, {Broguiere}, {Bronzwaer}, {Byun}, {Carlstrom}, {Chael}, {Chan}, {Chatterjee}, {Chatterjee}, {Chen}, {Chen}, {Cho}, {Christian}, {Conway}, {Cordes}, {Crew}, {Cui}, {Davelaar}, {De Laurentis}, {Deane}, {Dempsey}, {Desvignes}, {Dexter}, {Doeleman}, {Eatough}, {Falcke}, {Fish}, {Fomalont}, {Fraga-Encinas}, {Freeman}, {Friberg}, {Fromm}, {G{\'o}mez}, {Galison}, {Gammie}, {Garc{\'\i}a}, {Gentaz}, {Georgiev}, {Goddi}, {Gold}, {Gu}, {Gurwell}, {Hada}, {Hecht}, {Hesper}, {Ho}, {Ho}, {Honma}, {Huang}, {Huang}, {Hughes}, {Ikeda}, {Inoue}, {Issaoun}, {James}, {Jannuzi}, {Janssen}, {Jeter}, {Jiang}, {Johnson}, {Jorstad}, {Jung}, {Karami}, {Karuppusamy}, {Kawashima}, {Keating}, {Kettenis}, {Kim}, {Kim}, {Kim}, {Kino}, {Koay}, {Koch}, {Koyama}, {Kramer}, {Kramer}, {Krichbaum}, {Kuo}, {Lauer}, {Lee}, {Li}, {Li}, {Lindqvist}, {Liu}, {Liuzzo}, {Lo}, {Lobanov}, {Loinard}, {Lonsdale}, {Lu}, {MacDonald}, {Mao}, {Markoff}, {Marrone}, {Marscher}, {Mart{\'\i}-Vidal}, {Matsushita}, {Matthews}, {Medeiros}, {Menten}, {Mizuno}, {Mizuno}, {Moran}, {Moriyama}, {Moscibrodzka}, {M{\"u}ller}, {Nagai}, {Nagar}, {Nakamura}, {Narayan}, {Narayanan}, {Natarajan}, {Neri}, {Ni}, {Noutsos}, {Okino}, {Olivares}, {Ortiz-Le{\'o}n}, {Oyama}, {{\"O}zel}, {Palumbo}, {Patel}, {Pen}, {Pesce}, {Pi{\'e}tu}, {Plambeck}, {PopStefanija}, {Porth}, {Prather}, {Preciado-L{\'o}pez}, {Psaltis}, {Pu}, {Ramakrishnan}, {Rao}, {Rawlings}, {Raymond}, {Rezzolla}, {Ripperda}, {Roelofs}, {Rogers}, {Ros}, {Rose}, {Roshanineshat}, {Rottmann}, {Roy}, {Ruszczyk}, {Ryan}, {Rygl}, {S{\'a}nchez}, {S{\'a}nchez-Arguelles}, {Sasada}, {Savolainen}, {Schloerb}, {Schuster}, {Shao}, {Shen}, {Small}, {Sohn}, {SooHoo}, {Tazaki}, {Tiede}, {Tilanus}, {Titus}, {Toma}, {Torne}, {Trent}, {Trippe}, {Tsuda}, {van Bemmel}, {van Langevelde}, {van Rossum}, {Wagner}, {Wardle}, {Weintroub}, {Wex}, {Wharton}, {Wielgus}, {Wong}, {Wu}, {Young}, \& {Young}}]{EHT2019}
{Event Horizon Telescope Collaboration}, {Akiyama}, K., {Alberdi}, A., {et~al.} 2019, \apjl, 875, L1

\bibitem[{{Event Horizon Telescope Collaboration} {et~al.}(2021{\natexlab{a}}){Event Horizon Telescope Collaboration}, {Akiyama}, {Algaba}, {Alberdi}, {Alef}, {Anantua}, {Asada}, {Azulay}, {Baczko}, {Ball}, {Balokovi{\'c}}, {Barrett}, {Benson}, {Bintley}, {Blackburn}, {Blundell}, {Boland}, {Bouman}, {Bower}, {Boyce}, {Bremer}, {Brinkerink}, {Brissenden}, {Britzen}, {Broderick}, {Broguiere}, {Bronzwaer}, {Byun}, {Carlstrom}, {Chael}, {Chan}, {Chatterjee}, {Chatterjee}, {Chen}, {Chen}, {Chesler}, {Cho}, {Christian}, {Conway}, {Cordes}, {Crawford}, {Crew}, {Cruz-Osorio}, {Cui}, {Davelaar}, {De Laurentis}, {Deane}, {Dempsey}, {Desvignes}, {Dexter}, {Doeleman}, {Eatough}, {Falcke}, {Farah}, {Fish}, {Fomalont}, {Ford}, {Fraga-Encinas}, {Freeman}, {Friberg}, {Fromm}, {Fuentes}, {Galison}, {Gammie}, {Garc{\'\i}a}, {Gentaz}, {Georgiev}, {Goddi}, {Gold}, {G{\'o}mez}, {G{\'o}mez-Ruiz}, {Gu}, {Gurwell}, {Hada}, {Haggard}, {Hecht}, {Hesper}, {Ho}, {Ho}, {Honma}, {Huang}, {Huang}, {Hughes}, {Ikeda}, {Inoue}, {Issaoun}, {James}, {Jannuzi}, {Janssen}, {Jeter}, {Jiang}, {Jimenez-Rosales}, {Johnson}, {Jorstad}, {Jung}, {Karami}, {Karuppusamy}, {Kawashima}, {Keating}, {Kettenis}, {Kim}, {Kim}, {Kim}, {Kim}, {Kino}, {Koay}, {Kofuji}, {Koch}, {Koyama}, {Kramer}, {Kramer}, {Krichbaum}, {Kuo}, {Lauer}, {Lee}, {Levis}, {Li}, {Li}, {Lindqvist}, {Lico}, {Lindahl}, {Liu}, {Liu}, {Liuzzo}, {Lo}, {Lobanov}, {Loinard}, {Lonsdale}, {Lu}, {MacDonald}, {Mao}, {Marchili}, {Markoff}, {Marrone}, {Marscher}, {Mart{\'\i}-Vidal}, {Matsushita}, {Matthews}, {Medeiros}, {Menten}, {Mizuno}, {Mizuno}, {Moran}, {Moriyama}, {Moscibrodzka}, {M{\"u}ller}, {Musoke}, {Mej{\'\i}as}, {Michalik}, {Nadolski}, {Nagai}, {Nagar}, {Nakamura}, {Narayan}, {Narayanan}, {Natarajan}, {Nathanail}, {Neilsen}, {Neri}, {Ni}, {Noutsos}, {Nowak}, {Okino}, {Olivares}, {Ortiz-Le{\'o}n}, {Oyama}, {{\"O}zel}, {Palumbo}, {Park}, {Patel}, {Pen}, {Pesce}, {Pi{\'e}tu}, {Plambeck}, {PopStefanija}, {Porth}, {P{\"o}tzl}, {Prather}, {Preciado-L{\'o}pez}, {Psaltis}, {Pu}, {Ramakrishnan}, {Rao}, {Rawlings}, {Raymond}, {Rezzolla}, {Ricarte}, {Ripperda}, {Roelofs}, {Rogers}, {Ros}, {Rose}, {Roshanineshat}, {Rottmann}, {Roy}, {Ruszczyk}, {Rygl}, {S{\'a}nchez}, {S{\'a}nchez-Arguelles}, \& {Sasada}}]{EventHorizonTelescope2021a}
{Event Horizon Telescope Collaboration}, {Akiyama}, K., {Algaba}, J.~C., {et~al.} 2021{\natexlab{a}}, \apjl, 910, L12

\bibitem[{{Event Horizon Telescope Collaboration} {et~al.}(2021{\natexlab{b}}){Event Horizon Telescope Collaboration}, {Akiyama}, {Algaba}, {Alberdi}, {Alef}, {Anantua}, {Asada}, {Azulay}, {Baczko}, {Ball}, {Balokovi{\'c}}, {Barrett}, {Benson}, {Bintley}, {Blackburn}, {Blundell}, {Boland}, {Bouman}, {Bower}, {Boyce}, {Bremer}, {Brinkerink}, {Brissenden}, {Britzen}, {Broderick}, {Broguiere}, {Bronzwaer}, {Byun}, {Carlstrom}, {Chael}, {Chan}, {Chatterjee}, {Chatterjee}, {Chen}, {Chen}, {Chesler}, {Cho}, {Christian}, {Conway}, {Cordes}, {Crawford}, {Crew}, {Cruz-Osorio}, {Cui}, {Davelaar}, {De Laurentis}, {Deane}, {Dempsey}, {Desvignes}, {Dexter}, {Doeleman}, {Eatough}, {Falcke}, {Farah}, {Fish}, {Fomalont}, {Ford}, {Fraga-Encinas}, {Friberg}, {Fromm}, {Fuentes}, {Galison}, {Gammie}, {Garc{\'\i}a}, {Gelles}, {Gentaz}, {Georgiev}, {Goddi}, {Gold}, {G{\'o}mez}, {G{\'o}mez-Ruiz}, {Gu}, {Gurwell}, {Hada}, {Haggard}, {Hecht}, {Hesper}, {Himwich}, {Ho}, {Ho}, {Honma}, {Huang}, {Huang}, {Hughes}, {Ikeda}, {Inoue}, {Issaoun}, {James}, {Jannuzi}, {Janssen}, {Jeter}, {Jiang}, {Jimenez-Rosales}, {Johnson}, {Jorstad}, {Jung}, {Karami}, {Karuppusamy}, {Kawashima}, {Keating}, {Kettenis}, {Kim}, {Kim}, {Kim}, {Kim}, {Kino}, {Koay}, {Kofuji}, {Koch}, {Koyama}, {Kramer}, {Kramer}, {Krichbaum}, {Kuo}, {Lauer}, {Lee}, {Levis}, {Li}, {Li}, {Lindqvist}, {Lico}, {Lindahl}, {Liu}, {Liu}, {Liuzzo}, {Lo}, {Lobanov}, {Loinard}, {Lonsdale}, {Lu}, {MacDonald}, {Mao}, {Marchili}, {Markoff}, {Marrone}, {Marscher}, {Mart{\'\i}-Vidal}, {Matsushita}, {Matthews}, {Medeiros}, {Menten}, {Mizuno}, {Mizuno}, {Moran}, {Moriyama}, {Moscibrodzka}, {M{\"u}ller}, {Musoke}, {Mus Mej{\'\i}as}, {Michalik}, {Nadolski}, {Nagai}, {Nagar}, {Nakamura}, {Narayan}, {Narayanan}, {Natarajan}, {Nathanail}, {Neilsen}, {Neri}, {Ni}, {Noutsos}, {Nowak}, {Okino}, {Olivares}, {Ortiz-Le{\'o}n}, {Oyama}, {{\"O}zel}, {Palumbo}, {Park}, {Patel}, {Pen}, {Pesce}, {Pi{\'e}tu}, {Plambeck}, {PopStefanija}, {Porth}, {P{\"o}tzl}, {Prather}, {Preciado-L{\'o}pez}, {Psaltis}, {Pu}, {Ramakrishnan}, {Rao}, {Rawlings}, {Raymond}, {Rezzolla}, {Ricarte}, {Ripperda}, {Roelofs}, {Rogers}, {Ros}, {Rose}, {Roshanineshat}, {Rottmann}, {Roy}, {Ruszczyk}, {Rygl}, {S{\'a}nchez}, \& {S{\'a}nchez-Arguelles}}]{EventHorizonTelescope2021b}
{Event Horizon Telescope Collaboration}, {Akiyama}, K., {Algaba}, J.~C., {et~al.} 2021{\natexlab{b}}, \apjl, 910, L13

\bibitem[{{Everitt} {et~al.}(2011){Everitt}, {Debra}, {Parkinson}, {Turneaure}, {Conklin}, {Heifetz}, {Keiser}, {Silbergleit}, {Holmes}, {Kolodziejczak}, {Al-Meshari}, {Mester}, {Muhlfelder}, {Solomonik}, {Stahl}, {Worden}, {Bencze}, {Buchman}, {Clarke}, {Al-Jadaan}, {Al-Jibreen}, {Li}, {Lipa}, {Lockhart}, {Al-Suwaidan}, {Taber}, \& {Wang}}]{Everitt2011}
{Everitt}, C.~W.~F., {Debra}, D.~B., {Parkinson}, B.~W., {et~al.} 2011, \prl, 106, 221101

\bibitem[{{Fern{\'a}ndez Fern{\'a}ndez} {et~al.}(2023){Fern{\'a}ndez Fern{\'a}ndez}, {Della Monica}, \& {de Martino}}]{Fernandez2023}
{Fern{\'a}ndez Fern{\'a}ndez}, R., {Della Monica}, R., \& {de Martino}, I. 2023, \jcap, 2023, 039

\bibitem[{{Genzel} {et~al.}(2010){Genzel}, {Eisenhauer}, \& {Gillessen}}]{Genzel2010}
{Genzel}, R., {Eisenhauer}, F., \& {Gillessen}, S. 2010, Reviews of Modern Physics, 82, 3121

\bibitem[{{Ghez} {et~al.}(1998){Ghez}, {Klein}, {Morris}, \& {Becklin}}]{Ghez1998}
{Ghez}, A.~M., {Klein}, B.~L., {Morris}, M., \& {Becklin}, E.~E. 1998, \apj, 509, 678

\bibitem[{{Gillessen} {et~al.}(2017){Gillessen}, {Plewa}, {Eisenhauer}, {Sari}, {Waisberg}, {Habibi}, {Pfuhl}, {George}, {Dexter}, {von Fellenberg}, {Ott}, \& {Genzel}}]{Gillessen2017}
{Gillessen}, S., {Plewa}, P.~M., {Eisenhauer}, F., {et~al.} 2017, \apj, 837, 30

\bibitem[{{Gravity+ Collaboration} {et~al.}(2022){Gravity+ Collaboration}, {Abuter}, {Alarcon}, {Allouche}, {Amorim}, {Bailet}, {Bedigan}, {Berdeu}, {Berger}, {Berio}, {Bigioli}, {Blaho}, {Boebion}, {Bolzer}, {Bonnet}, {Bourdarot}, {Bourget}, {Brandner}, {Cardenas}, {Conzelmann}, {Comin}, {Cl{\'e}net}, {Courtney-Barrer}, {Dallilar}, {Davies}, {Defr{\`e}re}, {Delboulb{\'e}}, {Delplancke-Str{\"o}bele}, {Dembet}, {de Zeeuw}, {Drescher}, {Eckart}, {{\'E}douard}, {Eisenhauer}, {Fabricius}, {Feuchtgruber}, {Finger}, {F{\"o}rster Schreiber}, {Fuenteseca}, {Garcia}, {Garcia}, {Gao}, {Gendron}, {Genzel}, {Gil}, {Gillessen}, {Gomes}, {Gont{\'e}}, {Gouvret}, {Guajardo}, {Guidolin}, {Guieu}, {Guzmann}, {Hackenberg}, {Haddad}, {Hartl}, {Haubois}, {Hau{\ss}mann}, {Hei{\ss}el}, {Henning}, {Hippler}, {H{\"o}nig}, {Horrobin}, {Hubin}, {Jacqmart}, {Jocou}, {Kaufer}, {Kervella}, {Kirchbauer}, {Kolb}, {Korhonen}, {Kreidberg}, {Krempl}, {Lacour}, {Lagarde}, {Lai}, {Lapeyr{\`e}re}, {Laugier}, {Le Bouquin}, {Leftley}, {L{\'e}na}, {Lewis}, {Lutz}, {Magnard}, {Mang}, {Marcotto}, {Maurel}, {M{\'e}rand}, {Millour}, {More}, {Nowacki}, {Nowak}, {Oberti}, {Olivares}, {Ott}, {Pallanca}, {Paumard}, {Perraut}, {Perrin}, {Petrov}, {Pfuhl}, {Pourr{\'e}}, {Rabien}, {Rau}, {Riquelme}, {Robbe-Dubois}, {Rochat}, {Salman}, {Scherbarth}, {Sch{\"o}ller}, {Schubert}, {Schuhler}, {Shangguan}, {Shchekaturov}, {Shimizu}, {Scheithauer}, {Sevin}, {Soenke}, {Soulez}, {Spang}, {Stadler}, {Straubmeier}, {Sturm}, {Sykes}, {Tacconi}, {Tischer}, {Tristram}, {Vincent}, {von Fellenberg}, {Uysal}, {Widmann}, {Wieprecht}, {Wiezorrek}, {Woillez}, {Yaz{\i}c{\i}}, \& {Zins}}]{GravityPlus2022}
{Gravity+ Collaboration}, {Abuter}, R., {Alarcon}, P., {et~al.} 2022, The Messenger, 189, 17

\bibitem[{{Gravity Collaboration} {et~al.}(2018){Gravity Collaboration}, {Abuter}, {Amorim}, {Anugu}, {Baub{\"o}ck}, {Benisty}, {Berger}, {Blind}, {Bonnet}, {Brandner}, {Buron}, {Collin}, {Chapron}, {Cl{\'e}net}, {Coud{\'e} Du Foresto}, {de Zeeuw}, {Deen}, {Delplancke-Str{\"o}bele}, {Dembet}, {Dexter}, {Duvert}, {Eckart}, {Eisenhauer}, {Finger}, {F{\"o}rster Schreiber}, {F{\'e}dou}, {García}, {García Lopez}, {Gao}, {Gendron}, {Genzel}, {Gillessen}, {Gordo}, {Habibi}, {Haubois}, {Haug}, {Hau{\ss}mann}, {Henning}, {Hippler}, {Horrobin}, {Hubert}, {Hubin}, {Jimenez Rosales}, {Jochum}, {Jocou}, {Kaufer}, {Kellner}, {Kendrew}, {Kervella}, {Kok}, {Kulas}, {Lacour}, {Lapeyr{\`e}re}, {Lazareff}, {Le Bouquin}, {L{\'e}na}, {Lippa}, {Lenzen}, {M{\'e}rand}, {M{\"u}ler}, {Neumann}, {Ott}, {Palanca}, {Paumard}, {Pasquini}, {Perraut}, {Perrin}, {Pfuhl}, {Plewa}, {Rabien}, {Ram{\'\i}rez}, {Ramos}, {Rau}, {Rodr{\'\i}guez-Coira}, {Rohloff}, {Rousset}, {Sanchez-Bermudez}, {Scheithauer}, {Sch{\"o}ller}, {Schuler}, {Spyromilio}, {Straub}, {Straubmeier}, {Sturm}, {Tacconi}, {Tristram}, {Vincent}, {von Fellenberg}, {Wank}, {Waisberg}, {Widmann}, {Wieprecht}, {Wiest}, {Wiezorrek}, {Woillez}, {Yazici}, {Ziegler}, \& {Zins}}]{GravityCollaboration2018}
{Gravity Collaboration}, {Abuter}, R., {Amorim}, A., {et~al.} 2018, \aap, 615, L15

\bibitem[{{Gravity Collaboration} {et~al.}(2020){Gravity Collaboration}, {Abuter}, {Amorim}, {Baub{\"o}ck}, {Berger}, {Bonnet}, {Brandner}, {Cardoso}, {Cl{\'e}net}, {de Zeeuw}, {Dexter}, {Eckart}, {Eisenhauer}, {F{\"o}rster Schreiber}, {García}, {Gao}, {Gendron}, {Genzel}, {Gillessen}, {Habibi}, {Haubois}, {Henning}, {Hippler}, {Horrobin}, {Jim{\'e}nez-Rosales}, {Jochum}, {Jocou}, {Kaufer}, {Kervella}, {Lacour}, {Lapeyr{\`e}re}, {Le Bouquin}, {L{\'e}na}, {Nowak}, {Ott}, {Paumard}, {Perraut}, {Perrin}, {Pfuhl}, {Rodr{\'\i}guez-Coira}, {Shangguan}, {Scheithauer}, {Stadler}, {Straub}, {Straubmeier}, {Sturm}, {Tacconi}, {Vincent}, {von Fellenberg}, {Waisberg}, {Widmann}, {Wieprecht}, {Wiezorrek}, {Woillez}, {Yazici}, \& {Zins}}]{GravityCollaboration2020}
{Gravity Collaboration}, {Abuter}, R., {Amorim}, A., {et~al.} 2020, \aap, 636, L5

\bibitem[{{Hees} {et~al.}(2017){Hees}, {Do}, {Ghez}, {Martinez}, {Naoz}, {Becklin}, {Boehle}, {Chappell}, {Chu}, {Dehghanfar}, {Kosmo}, {Lu}, {Matthews}, {Morris}, {Sakai}, {Sch{\"o}del}, \& {Witzel}}]{Hees2017}
{Hees}, A., {Do}, T., {Ghez}, A.~M., {et~al.} 2017, Physical Review Letters, 118, 211101

\bibitem[{{Johnson} {et~al.}(2023){Johnson}, {Akiyama}, {Blackburn}, {Bouman}, {Broderick}, {Cardoso}, {Fender}, {Fromm}, {Galison}, {G{\'o}mez}, {Haggard}, {Lister}, {Lobanov}, {Markoff}, {Narayan}, {Natarajan}, {Nichols}, {Pesce}, {Younsi}, {Chael}, {Chatterjee}, {Chaves}, {Doboszewski}, {Dodson}, {Doeleman}, {Elder}, {Fitzpatrick}, {Haworth}, {Houston}, {Issaoun}, {Kovalev}, {Levis}, {Lico}, {Marcoci}, {Martens}, {Nagar}, {Oppenheimer}, {Palumbo}, {Ricarte}, {Rioja}, {Roelofs}, {Thresher}, {Tiede}, {Weintroub}, \& {Wielgus}}]{Johnson2023}
{Johnson}, M.~D., {Akiyama}, K., {Blackburn}, L., {et~al.} 2023, Galaxies, 11, 61

\bibitem[{{Keane} {et~al.}(2015){Keane}, {Bhattacharyya}, {Kramer}, {Stappers}, {Keane}, {Bhattacharyya}, {Kramer}, {Stappers}, {Bates}, {Burgay}, {Chatterjee}, {Champion}, {Eatough}, {Hessels}, {Janssen}, {Lee}, {van Leeuwen}, {Margueron}, {Oertel}, {Possenti}, {Ransom}, {Theureau}, \& {Torne}}]{Keane2015}
{Keane}, E., {Bhattacharyya}, B., {Kramer}, M., {et~al.} 2015, in Advancing Astrophysics with the Square Kilometre Array (AASKA14), 40

\bibitem[{{Luminet}(1979)}]{Luminet1979}
{Luminet}, J.~P. 1979, \aap, 75, 228

\bibitem[{Meurer {et~al.}(2017)Meurer, Smith, Paprocki, \v{C}ert\'{i}k, Kirpichev, Rocklin, Kumar, Ivanov, Moore, Singh, Rathnayake, Vig, Granger, Muller, Bonazzi, Gupta, Vats, Johansson, Pedregosa, Curry, Terrel, Rou\v{c}ka, Saboo, Fernando, Kulal, Cimrman, \& Scopatz}]{Meurer2017}
Meurer, A., Smith, C.~P., Paprocki, M., {et~al.} 2017, PeerJ Computer Science, 3, e103

\bibitem[{{Nan} {et~al.}(2011){Nan}, {Li}, {Jin}, {Wang}, {Zhu}, {Zhu}, {Zhang}, {Yue}, \& {Qian}}]{Nan2011}
{Nan}, R., {Li}, D., {Jin}, C., {et~al.} 2011, International Journal of Modern Physics D, 20, 989

\bibitem[{{Poisson} \& {Will}(2014)}]{Poisson2014}
{Poisson}, E. \& {Will}, C.~M. 2014, {Gravity}

\bibitem[{{Pound} \& {Rebka}(1960)}]{Pound1960}
{Pound}, R.~V. \& {Rebka}, G.~A. 1960, \prl, 4, 274

\bibitem[{{Press} {et~al.}(1992){Press}, {Teukolsky}, {Vetterling}, \& {Flannery}}]{Press1992}
{Press}, W.~H., {Teukolsky}, S.~A., {Vetterling}, W.~T., \& {Flannery}, B.~P. 1992, {Numerical recipes in C. The art of scientific computing}

\bibitem[{{Psaltis} {et~al.}(2020){Psaltis}, {Medeiros}, {Christian}, {{\"O}zel}, {Akiyama}, {Alberdi}, {Alef}, {Asada}, {Azulay}, {Ball}, {Balokovi{\'c}}, {Barrett}, {Bintley}, {Blackburn}, {Boland}, {Bower}, {Bremer}, {Brinkerink}, {Brissenden}, {Britzen}, {Broguiere}, {Bronzwaer}, {Byun}, {Carlstrom}, {Chael}, {Chan}, {Chatterjee}, {Chatterjee}, {Chen}, {Chen}, {Cho}, {Conway}, {Cordes}, {Crew}, {Cui}, {Davelaar}, {De Laurentis}, {Deane}, {Dempsey}, {Desvignes}, {Dexter}, {Eatough}, {Falcke}, {Fish}, {Fomalont}, {Fraga-Encinas}, {Friberg}, {Fromm}, {Gammie}, {Garc{\'\i}a}, {Gentaz}, {Goddi}, {G{\'o}mez}, {Gu}, {Gurwell}, {Hada}, {Hesper}, {Ho}, {Ho}, {Honma}, {Huang}, {Huang}, {Hughes}, {Inoue}, {Issaoun}, {James}, {Jannuzi}, {Janssen}, {Jiang}, {Jimenez-Rosales}, {Johnson}, {Jorstad}, {Jung}, {Karami}, {Karuppusamy}, {Kawashima}, {Keating}, {Kettenis}, {Kim}, {Kim}, {Kim}, {Kino}, {Koay}, {Koch}, {Koyama}, {Kramer}, {Kramer}, {Krichbaum}, {Kuo}, {Lauer}, {Lee}, {Li}, {Li}, {Lindqvist}, {Lico}, {Liu}, {Liu}, {Liuzzo}, {Lo}, {Lobanov}, {Lonsdale}, {Lu}, {Mao}, {Markoff}, {Marrone}, {Marscher}, {Mart{\'\i}-Vidal}, {Matsushita}, {Mizuno}, {Mizuno}, {Moran}, {Moriyama}, {Moscibrodzka}, {M{\"u}ller}, {Musoke}, {Mus Mej{\'\i}as}, {Nagai}, {Nagar}, {Narayan}, {Narayanan}, {Natarajan}, {Neri}, {Noutsos}, {Okino}, {Olivares}, {Oyama}, {Palumbo}, {Park}, {Patel}, {Pen}, {Pi{\'e}tu}, {Plambeck}, {PopStefanija}, {Prather}, {Preciado-L{\'o}pez}, {Ramakrishnan}, {Rao}, {Rawlings}, {Raymond}, {Ripperda}, {Roelofs}, {Rogers}, {Ros}, {Rose}, {Roshanineshat}, {Rottmann}, {Roy}, {Ruszczyk}, {Ryan}, {Rygl}, {S{\'a}nchez}, {S{\'a}nchez-Arguelles}, {Sasada}, {Savolainen}, {Schloerb}, {Schuster}, {Shao}, {Shen}, {Small}, {Sohn}, {SooHoo}, {Tazaki}, {Tilanus}, {Titus}, {Torne}, {Trent}, {Traianou}, {Trippe}, {van Bemmel}, {van Langevelde}, {van Rossum}, {Wagner}, {Wardle}, {Ward-Thompson}, {Weintroub}, {Wex}, {Wharton}, {Wielgus}, {Wong}, {Wu}, {Yoon}, {Young}, {Young}, {Younsi}, {Yuan}, {Yuan}, {Zhao}, \& {EHT Collaboration}}]{Psaltis2020}
{Psaltis}, D., {Medeiros}, L., {Christian}, P., {et~al.} 2020, \prl, 125, 141104

\bibitem[{{Skidmore} {et~al.}(2015){Skidmore}, {TMT International Science Development Teams}, \& {Science Advisory Committee}}]{Skidmore2015}
{Skidmore}, W., {TMT International Science Development Teams}, \& {Science Advisory Committee}, T. 2015, Research in Astronomy and Astrophysics, 15, 1945

\bibitem[{Straumann(2013)}]{Straumann2013}
Straumann, N. 2013, {General Relativity}, Graduate Texts in Physics (Dordrecht: Springer)

\bibitem[{Sturm {et~al.}(2024)Sturm, Davies, Alves, Cl{\'e}net, Kotilainen, Monna, Nicklas, Pott, Tolstoy, Vulcani, Achren, Annadevara, Anwand-Heerwart, Arcidiacono, Barboza, Barl, Baudoz, Bender, Bezawada, Biondi, Bizenberger, Blin, Bon{\'e}, Bonifacio, Borgo, van~den Born, Buey, Cao, Chapron, Chauvin, Chemla, Cloiseau, Cohen, Colin, Czoske, Dette, Deysenroth, Dijkstra, Dreizler, Dupuis, van Egmond, Eisenhauer, Elswijk, Emslander, Fabricius, Fasola, Ferreira, Schreiber, Fontana, Gaudemard, Gautherot, Gendron, Gennet, Genzel, Ghouchou, Gillessen, Gratadour, Grazian, Grupp, Guieu, Gullieuszik, de~Haan, Hartke, Hartl, Haussmann, Helin, Hess, Hofferbert, Huber, Huby, Huet, Ives, Janssen, Jaufmann, Jilg, Jodlbauer, Jost, Kausch, Kellermann, Kerber, Kravcar, Kravchenko, Kulcs{\'a}r, Kuncarayakti, Kunst, Kwast, Lang, Lange, Lapeyrere, Ruyet, Leschinski, Locatelli, Massari, Mattila, Mei, Merlin, Meyer, Michel, Mohr, Montarg{\`e}s, M{\"u}ller, M{\"u}nch, Navarro, Neumann, Neumayer, Neumeier, Pedichini, Pfl{\"u}ger, Piazzesi, Pinard, Porras, Portulari, Przybilla, Rabien, Raffard, Ragazzoni, Ramlau, Ramos, Ramsay, Raynaud, Rhode, Richter, Rix, Rodenhuis, Rohloff, Romp, Rousselot, Sabha, Sassolas, Schlichter, Schuil, Schweitzer, Seemann, Sevin, Simioni, Spallek, S{\"o}nmez, Suuronen, Taburet, Thomas, Tisserand, Vaccari, Valenti, Kleijn, Verdugo, Vidal, Wagner, Wegner, van Winden, Witschel, Zanella, Zeilinger, Ziegleder, \& Ziegler}]{Sturm2024}
Sturm, E., Davies, R., Alves, J., {et~al.} 2024, in Ground-based and Airborne Instrumentation for Astronomy X, ed. J.~J. Bryant, K.~Motohara, \& J.~R.~D. Vernet, Vol. 13096, International Society for Optics and Photonics (SPIE), 1309611

\bibitem[{{Taff}(1985)}]{Taff1985}
{Taff}, L.~G. 1985, {Celestial mechanics : a computational guide for the practitioner}

\bibitem[{{Vessot} {et~al.}(1980){Vessot}, {Levine}, {Mattison}, {Blomberg}, {Hoffman}, {Nystrom}, {Farrel}, {Decher}, {Eby}, \& {Baugher}}]{Vessot1980}
{Vessot}, R.~F.~C., {Levine}, M.~W., {Mattison}, E.~M., {et~al.} 1980, \prl, 45, 2081

\bibitem[{{Vincent} {et~al.}(2011){Vincent}, {Paumard}, {Gourgoulhon}, \& {Perrin}}]{Vincent2011}
{Vincent}, F.~H., {Paumard}, T., {Gourgoulhon}, E., \& {Perrin}, G. 2011, Classical and Quantum Gravity, 28, 225011

\bibitem[{{Webster} \& {Murdin}(1972)}]{Webster1972}
{Webster}, B.~L. \& {Murdin}, P. 1972, \nat, 235, 37

\bibitem[{{Will}(2014)}]{Will2014}
{Will}, C.~M. 2014, Living Reviews in Relativity, 17, 4

\end{thebibliography}

  \end{document}